\documentclass[english,12pt,onecolumn]{IEEEtran}

\usepackage{babel}
\usepackage{graphicx}
\usepackage{subfigure}
\usepackage{float}
\usepackage{tabularx}
\usepackage{epstopdf}
\usepackage{url}
\usepackage{marvosym}
\usepackage{amssymb}
\usepackage{amsthm}
\usepackage{tikz}
\usepackage{textcomp}
\usepackage{marvosym}
\usepackage{booktabs}
\usepackage{ulem}
\usepackage[cmex10]{amsmath}
\usepackage{amsmath, bm}

\newtheorem{theorem}{Theorem}

\newtheorem{assumption}{Assumption}

\newtheorem{remark}{Remark}
\newtheorem{proposition}{Proposition}

\newcommand{\argmin}{\operatornamewithlimits{argmin}}

\DeclareMathOperator{\sgn}{sgn}

\DeclareMathOperator{\conv}{conv}

\begin{document}

\title{Safety-Critical Control of Euler-Lagrange Systems Subject to Multiple Obstacles and Velocity Constraints\thanks{This work was supported in part by the National Natural Science Foundation of China under Grants U1911401 and 61633007, and in part by the U.S. National Science Foundation under Grant EPCN-2210320.}}
\author{Zhi Liu\thanks{State Key Laboratory of Synthetical Automation for Process Industries, Northeastern University, Shenyang, 110004, China Email: {\tt lz170206@163.com}}, Si Wu\thanks{State Key Laboratory of Synthetical Automation for Process Industries, Northeastern University, Shenyang, 110004, China Email: {\tt wusixstx@163.com}}, Tengfei Liu\thanks{State Key Laboratory of Synthetical Automation for Process Industries, Northeastern University, Shenyang, 110004, China Email: {\tt tfliu@mail.neu.edu.cn}} and Zhong-Ping Jiang\thanks{Department of Electrical and Computer Engineering, New York University, Brooklyn, NY 11201, USA Email: {\tt zjiang@nyu.edu}}}
\date{\ }

\maketitle

\begin{abstract}
This paper studies the safety-critical control problem for Euler-Lagrange (EL) systems subject to multiple ball obstacles and velocity constraints in accordance with affordable velocity ranges. A key strategy is to exploit the underlying inner-outer-loop structure for the design of a new cascade controller for the class of EL systems. In particular, the outer-loop controller is developed based on quadratic programming (QP) to avoid ball obstacles and generate velocity reference signals fulfilling the velocity limitation. Taking full advantage of the conservation-of-energy property, a nonlinear velocity-tracking controller is designed to form the inner loop. One major difficulty is caused by the possible non-Lipschitz continuity of the standard QP algorithm when there are multiple constraints. To solve this problem, we propose a refined QP algorithm with the feasible set reshaped by an appropriately chosen positive basis such that the feasibility is retained while the resulting outer-loop controller is locally Lipschitz. It is proved that the constraint-satisfaction problem is solvable as long as the ball obstacles satisfy a mild distance condition. The proposed design is validated by numerical simulation and an experiment based on a $2$-link planar manipulator.
\end{abstract}


\section{Introduction}

Accomplishing complex tasks while ensuring safety is an essential capability of various engineering control systems, including vehicles, robotics, and industrial processes. The rapid development of computing, communication, and system integration techniques enables the embedded, real-time implementation of optimization-based controllers. Many recent safety-critical control results construct constraint-satisfaction-based reactive controllers such that the control signals fulfill safety requirements and are as close as possible to primary control commands generated by higher-level controllers or even human operators; see, e.g., \cite{Latombe-book-1991,Arkin-book-1998,Choset-book-2005,Ren-Beard-book-2008,Bullo-Cortes-Martinez-book-2009,Mesbahi-Egerstedt-book-2010,Lynch-Park-2017-Book} and \cite{Slotine-Siciliano-1991-ICAR}.

Among these results, (control) barrier functions play a central role in mathematically defining the state constraints and the corresponding feasible control inputs for control systems, motivating new control design techniques; see, e.g., \cite{Polak-Yang-Mayne-SIAMControl-1993,Wills-Heath-Auto-2004,Ngo-Mahony-Jiang-CDC-2005,Tee-Ge-Tay-Auto-2009,Wieland-Allgower-NOLCOS-2007,Prajna-Jadbabaie-Pappas-TAC-2007,Ames-Grizzle-Tabuada-CDC-2014,Wisniewski-Sloth-TAC-2016} and \cite{Romdlony-Jayawardhana-Auto-2016}. The substantial relationship between control Lyapunov functions given by \cite{Artstein-NA-1983} and control barrier function opens the door to a systematic development of a multi-objective control theory \cite{Ames-Xu-Grizzle-Tabuada-TAC-2017}. Indeed, for a control-affine system, an appropriately defined control barrier function entails linear inequality constraints on admissible control inputs for safety and more general purposes and allows computationally efficient integration of different control strategies. Zeroing barrier functions only assume an increasing property when the system state is outside the safe set \cite{Xu-Tabuada-Grizzle-Ames-IFAC-2015}. Variants of (control) barrier functions, including robust barrier functions \cite{Jankovic-Auto-2018} and high-order control barrier functions \cite{Xiao-TAC-2021}, have been developed to handle uncertainties and higher-order dynamics of the control system under consideration.

Quadratic programming (QP) is a powerful tool for real-time synthesis of controllers by incorporating different specifications \cite{Nakamura-1990-Book,Escande-Mansard-Wieber-IJRR-2014,Mellinger-Kumar-ICRA-2011,Ames-Powell-CPS-2013}. From a feasible control set, a QP algorithm selects the admissible control input closest to the primary control command signal. This way, constraints (defined by control barrier functions) and performance objectives (described by control Lyapunov functions) are naturally integrated within a unified framework \cite{Ames-Xu-Grizzle-Tabuada-TAC-2017}. The QP approach has also been extended to nonsmooth barrier functions \cite{Glotfelter-Paul-TAC-2021}. Applications include mobile robots \cite{Wang-Ames-Egerstedt-TRO-2017,Wilson-Egerstedt-CSM-2020}, robotic grasping \cite{Cortez-Oetomo-Manzie-Choong-TCST-2019,Singletary-Guffey-RAL-2022}, autonomous driving \cite{Ames-Xu-Grizzle-Tabuada-TAC-2017}, bipedal walking \cite{Xiong-Ames-RAL-2021}, exoskeletons \cite{Wu-Li-Kan-Gao-TC-2019}, human-machine interaction systems \cite{Singletary-Nilsson-Gurriet-Ames-IROS-2019}, etc. The techniques that safety-critical control uses to handle constraints are also promising to inspire new control strategies to solve relevant problems demanding real-time constraint satisfaction.

The Euler-Lagrange (EL) equation captures the motion of a large class of physical systems and has been widely used to study the safety control problem for mechanical systems. Developing a systematic solution to the safety-critical control of EL systems is highly demanded. With industrial robotic manipulators as a typical example, safety essentially requires that every point on the body keeps far away enough from cooperative and noncooperative objects. One obstacle point in the workspace usually results in obstacles in the configuration space (joint space). Compared with normal stabilizing and tracking controllers, a well-designed safety-critical controller should handle not only the nonlinear uncertain dynamics of the plant but also various safety constraints on real-time state of the plant, which cause significant technical difficulties. In previous results on safety control of EL systems, safety has been considered as a component of higher levels of control in hierarchical control systems which limits the EL system's real-time capabilities for fast and highly interactive operations in a cluttered environment; see e.g., \cite{Lozamo-TSMC-1981,Dietrich-RAM-2012} and \cite{Youakim-RAM-2017}. Artificial potential field methods couple the environment sensing feedback with the lowest level of the control but do not fully take into account the impact of the nonlinear uncertain dynamics of the plant \cite{Khatib-JRR-1986}. 

The QP algorithm and control barrier functions have been shown promising to solve the safety-critical control problems of EL systems. However, applying the QP algorithm to merge multiple barrier functions could result in a non-Lipschitz controller, which may cause chattering and affect the stability and durability of the control system \cite{Singletary-Guffey-RAL-2022}. Energy-based control barrier functions take advantage of the conservation-of-energy property of EL systems but only focus on single safety constraint \cite{Singletary-Kolathaya-CSL-2022}. Existing results on high-order barrier functions normally rely on known plant dynamics and possibly lead to infeasibility under tight safety constraints \cite{Xiao-TAC-2021}. Reference \cite{Cortez-Dimarogonas-Auto-2022} considers the case in which the states of the EL system are desired to be within generalized axis-aligned bounding boxes.

This paper takes a step forward toward solving the safety control problem for EL systems subject to multiple obstacles in the configuration space. Following the convention of constructive nonlinear control \cite{Krstic-Kanellakopoulos-Kokotovic-1995-Book}, we propose a controller featuring an inner-outer-loop structure. In particular, the outer-loop control law addresses position constraints generated by multiple obstacles with the generalized velocity as a virtual control input, and the inner-loop control law is devoted to velocity tracking.

It is well-known that necessary smoothness (at least local Lipschitz continuity) of the outer-loop control law is crucial for guaranteed feasibility and robustness in the presence of the inner-loop dynamics. However, designing a Lipschitz continuous control law is nontrivial when there are multiple safety constraints. Indeed, it is well-known that the solution to a QP problem may be non-Lipschitz if there are multiple constraints. Usually, the nonredundant active constraints do not readily satisfy the essential full-row-rank condition \cite{Hager-SIAMControl-1979} for the Lipschitz continuity of QP solutions. We mention that the control problem subject to multiple constraints has been studied in \cite{Wu-Liu-Niu-Jiang-RAL-2022,Wu-Liu-Egerstedt-Jiang-TAC-2023} and \cite{Cortez-CSL-2022}. However, these results rely on kinematic models, and do not fully take into account uncertain plant dynamics. 

This paper addresses the challenge by contributing a new class of QP-based control laws with a nested-loop structure. In particular, the constraints are projected onto an appropriately chosen positive basis such that the new, reshaped constraints retain the feasibility and still guarantee the satisfaction of the constraints. Moreover, under mild conditions, the new control law is Lipschitz continuous with respect to the generalized coordinates of the EL system. Then, taking advantage of the inherent conservation-of-energy property of EL systems, we design a nonlinear proportional control law for the inner loop for velocity tracking. The proposed design allows uncertain dynamics of the EL system. The advantage also lies in incorporating velocity limitations to meet practical requirements.

The layout of the paper is as follows. Section \ref{problem} presents the problem statement. Section \ref{section.design} gives a cascade controller design method for the safety control of EL systems. The resulting safety controller consists of a QP-based Lipschitz safety control law to handle multiple obstacles constraints, and a velocity tracking control law. Section \ref{T2} proposes a refined control law to reduce the computation load. Section \ref{section.simulation} employs a numerical simulation and an experiment of a 2-DOF manipulator to verify the methods proposed in Sections \ref{section.design} and \ref{T2}. Finally, some concluding remarks are offered in Section \ref{conclusion}.

\subsubsection*{Notations}

Most of the notations in this paper follow the convention of nonlinear control and can be found in \cite{Khalil-book-2002}. A continuous function $\alpha:\mathbb{R}\rightarrow\mathbb{R}$ is said to be of class $\mathcal{K}^e$ if it is strictly increasing and $\alpha(0)=0$. The matrix inequality $A<B$ for square matrices $A$ and $B$ means that the matrix $B-A$ is positive definite. For a function $f:\mathbb{R}^n\rightarrow\mathbb{R}$ differentiable at $x\in\mathbb{R}^n$, its gradient at $x\in\mathbb{R}^n$ is denoted as $\nabla f(x)$. The generalized directional derivative of a locally Lipschitz function $f:\mathbb{R}^n\rightarrow\mathbb{R}$ at $x\in\mathbb{R}^n$ with a vector $v\in\mathbb{R}^n$ is defined by
\begin{align}
	f^{\circ}(x,v)=\limsup_{y\rightarrow x,\lambda\downarrow 0}\frac{f(y+\lambda v)-f(y)}{\lambda}.\label{genralizedderivarive}
\end{align}
The generalized gradient of a locally Lipschitz function $f:\mathbb{R}^n\rightarrow\mathbb{R}$ at $x\in\mathbb{R}^n$ is defined by
\begin{align}
	\partial f(x)=\{\epsilon\in\mathbb{R}^n:f^{\circ}(x,v)\geq \epsilon^Tv,\forall v\in\mathbb{R}^n\}. \label{genralizedgradient}
\end{align}
For a vector-valued function $f:\mathbb{R}^m\rightarrow\mathbb{R}^n$, we use $f_{ci}:\mathbb{R}^m\rightarrow\mathbb{R},i\in\{1,\ldots,n\}$ to represent the $i$-th component of $f$. The symbol $\conv$ represents the convex hull. The sum of two sets $A_1$ and $A_2$ is defined as
\begin{align}
	A_1+A_2=\{a_1+a_2:a_1\in A_1,a_2\in A_2\}.\label{sumset}
\end{align}

\section{Problem Formulation}\label{problem}

Consider the model of an $n$-DOF EL system as follows:
\begin{align}
	M(q)\ddot{q}+C(q,\dot{q})\dot{q}+N(q,\dot{q})=u\label{EL}
\end{align}
where $q\in\mathbb{R}^n$ is the vector of generalized coordinates (positions), $M(q)\in\mathbb{R}^{n\times n}$ is the inertia matrix, $C(q,\dot{q})\dot{q}$ represents the Coriolis and centrifugal forces with $C(q,\dot{q})\in\mathbb{R}^{n\times n}$, $N(q,\dot{q})\in\mathbb{R}^n$ includes generalized resistance and gravity terms, and $u\in \mathbb{R}^n$ represents the vector of generalized forces. The following standing assumption is made on the EL system \cite{Murray-Li-Sastry-1994-book,Craig-book-2005}.
\begin{assumption}\label{assumption.EL}
	The system \eqref{EL} admits the following properties:
	\begin{itemize}
		\item Along the trajectories of the system \eqref{EL}, $\dot{M}-2C$ is a skew-symmetric matrix, that is,
		\begin{align}
			(\dot{M}-2C)+(\dot{M}-2C)^T=0;\label{M2C}
		\end{align}
		\item $M(q)$ is an invertible, symmetric matrix for any $q\in\mathbb{R}^n$ and there exist positive constants $\mu_{m1}$ and $\mu_{m2}$ such that
		\begin{align}
			\mu_{m1}I_{n\times n}< M(q)< \mu_{m2}I_{n\times n}\label{matrix}
		\end{align}
		for all $q\in\mathbb{R}^n$.
	\end{itemize}
\end{assumption}

In our problem setting, we consider $u$ as the control input. Given $v_c:\mathbb{R}_+\rightarrow\mathbb{R}^n$ as the velocity command signal, we focus on designing a controller for \eqref{EL} to satisfy constraints and minimally invade the velocity command signal $v_c$. In particular, throughout the control procedure, the generalized coordinates $q$ of the EL system are required to satisfy the following constraints:
\begin{align}
	h_i(q(t))\geq 0,\quad i=1,\ldots,N\label{safetyregion}
\end{align}
for all $t\geq 0$, with
\begin{align}
	h_i(q)=|q_{oi}-q|-d_{si},\label{safetyregionnot}
\end{align}
where $q_{oi}\in\mathbb{R}^n$ represents the obstacle positions that the generalized coordinates $q$ should keep far away from and $d_{si}$ is a positive constant representing the corresponding safety margin.

Moreover, the generalized velocities are required to stay within bounded ranges:
\begin{align}
	|\dot{q}(t)|\leq \bar{v}\label{barv}
\end{align}
for all $t\geq 0$, where $\bar{v}$ is a positive constant representing the velocity upper bound.

Without loss of generality, we assume that the velocity command signal $v_c$ is bounded and differentiable on the time-line.
\begin{assumption}\label{assumption.vc}
	There exist positive constants $\bar{v}_c$ and $v_c^d$ such that
	\begin{align}
		|v_c(t)|\leq \bar{v}_c,~
		|\dot{v}_c(t)|\leq v_c^d,\label{vclipschitz}
	\end{align}
	for all $t\geq 0$.
\end{assumption}

We also assume that the obstacles satisfy a mild distance condition.

\begin{assumption}\label{assumption.balldistance}
	There exist a constant $d_a>0$ satisfying $d_a<\min_{i=1,\ldots,N}\{d_{si}\}$, a constant $d_b>0$ and a constant $d_h>0$ satisfying $d_h<\min_{i=1,\ldots,N}\{d_{si}\}$ such that for any $q$ satisfying $h_i(q)\geq -d_h$ for all $i=1,\ldots,N$, if
	\begin{align}
		h_j(q)\leq \frac{d_b}{2},~~~~h_k(q)\leq\frac{d_b}{2}
	\end{align}
	hold for some $j,~k\in\{1,\ldots,N\}$, then
	\begin{align}
		|q_{oj}-q_{ok}|\leq d_a.\label{distancelimit}
	\end{align}
\end{assumption}

\begin{remark}
	Assumption \ref{assumption.balldistance} requires that the obstacles in the configuration space can be covered by several ball obstacles that are either intersecting or disjoint, and the positions of the intersecting ball obstacles are close enough.
\end{remark}

\begin{remark}
	Given a specific scenario of safety-critical control, one may first divide the configuration space into uniform-sized hypercubes and then assign balls with proper radius to the hypercubes to cover the unsafe region. Assumption \ref{assumption.balldistance} can be satisfied by choosing the hypercubes small enough and the balls large enough and adding balls according to the unsafe region. An example for safety-critical control of a $2$-link planar manipulator is given in Section \ref{section.simulation}.
\end{remark}

\section{Cascade Controller Design}
\label{section.design}

We follow the convention of cascade designs to solve the problem. In particular, we first consider the velocity $v$ as the virtual control input, and design a QP-based control law to incorporate the constraints on $q$ and $\dot{q}$ as well as the primary velocity command in Subsection \ref{section3a}. The idea of robust control barrier functions \cite{Xu-Tabuada-Grizzle-Ames-IFAC-2015} is used to guarantee the satisfaction of the constraints in the presence of velocity tracking errors. Moreover, we employ a positive basis to reshape the feasible set of the resulting QP problem such that feasibility is retained and the velocity reference signal $v^*$, i.e., the solution to the QP problem, is locally Lipschitz with respect to the generalized coordinates $q$. This makes it possible to further design a velocity controller with $u$ as the control input such that the velocity $\dot{q}$ tracks the ideal velocity reference signal $v^*$. Taking advantage of the inherent conservation-of-energy property of Euler-Lagrange systems, we design a nonlinear proportional control law for velocity-tracking control in Subsection \ref{section3b}. The stability analysis of the closed-loop system that integrates the outer-loop and inner-loop control laws is given in Subsection \ref{section3c}.

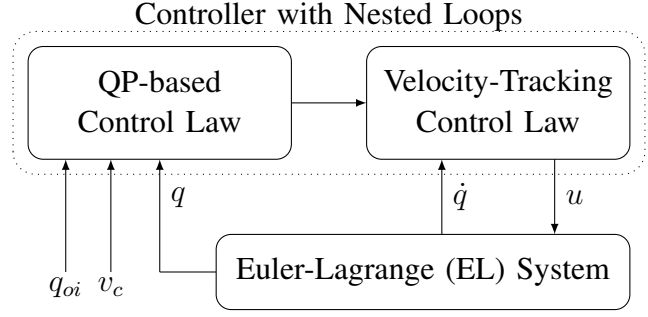
\begin{figure}[h!]
	\centering
	\begin{tikzpicture}[scale = 1,>=latex]
		\draw (40mm,19mm) node {Controller with Nested Loops};
		\draw [dotted,rounded corners=3mm] (-2mm,-2mm) rectangle (82mm,17mm);
		\draw [rounded corners=2mm] (0mm,0mm) rectangle (35mm,15mm);
		\draw (17.5mm,10mm) node {QP-based};
		\draw (17.5mm,5mm) node {Control Law};
		\draw [->] (35mm,7.5mm) -- (45mm,7.5mm);
		\draw [rounded corners=2mm] (45mm,0mm) rectangle (80mm,15mm);
		\draw (62.5mm,10mm) node {Velocity-Tracking};
		\draw (62.5mm,5mm) node {Control Law};
		\draw [rounded corners=2mm] (25mm,-20mm) rectangle (80mm,-10mm);
		\draw (52.5mm,-15mm) node {Euler-Lagrange (EL) System};
		\draw [->] (70mm,0mm) -- (70mm,-10mm);
		\draw (70mm,-5mm) node [right] {$u$};
		\draw [->] (55mm,-10mm) -- (55mm,0mm);
		\draw (55mm,-5mm) node [right] {$\dot{q}$};
		\draw (25mm,-15mm) -- (17.5mm,-15mm);
		\draw [->] (17.5mm,-15mm) -- (17.5mm,0mm);
		\draw (17.5mm,-5mm) node [right] {$q$};
		\draw [->] (5mm,-15mm) -- (5mm,0mm);
		\draw (5mm,-17mm) node {$q_{oi}$};

		\draw [->](1.1,-1.5) -- (1.1,0);
		
		\draw (1.1,-1.7)node {$v_c$};
	\end{tikzpicture}
	\caption{The proposed cascade control structure to handle constraints.}
	\label{figure.cascadecontrolstructure}
\end{figure}

\subsection{QP-based Control Law to Handle Position Constraints}
\label{section3a}

The problem formulation intuitively motivates a QP-based integration of the safety constraints and the primary velocity command (\cite{Ames-Xu-Grizzle-Tabuada-TAC-2017}):
\begin{align}
	&v^*=\mathop{\argmin}\limits_{v^*\in\mathbb{R}^n}|v^*-v_c|^2\label{originQP1}\\
	&\text{s.t.}~\frac{(q_{oi}-q)^T}{|q_{oi}-q|}v^*\leq\alpha_p(|q_{oi}-q|-d_{si}),~ i=1,\ldots,N\label{originQP2}
\end{align}
where $\alpha_p$ is a class $\mathcal{K}^e$ function. To handle the velocity-tracking error $\dot{q}-v^*$, one would introduce an additional robust margin $d_r$ to the position constraints. Then, the constraint \eqref{originQP2} should be modified as
\begin{align}
	\frac{(q_{oi}-q)^T}{|q_{oi}-q|}v^*\leq\alpha_p(|q_{oi}-q|-d_{si})-d_r,\quad i=1,\ldots,N.\label{robustsafetyconstraint}
\end{align}
However, a QP-based control law may not be able to ensure the Lipschitz continuity of the solution $v^*$ when there are more than one constraint (i.e., $N\geq 2$; see \cite{Hager-SIAMControl-1979}). This means that the QP algorithm above may not be readily applicable as an outer-loop control law for the intended constructive design.

We contribute a refined feasible-set reshaping technique to solve this problem. In particular, under the condition that $q\neq q_{o1},\ldots,q_{oN}$, we employ a positive basis to represent the constraints, and propose the following refined QP-based control law:
\begin{align}
	& v^*=\mathop{\argmin}\limits_{v^*\in\mathbb{R}^n}|v^*-v_c|^2\label{QPvirtualcontrolllaw1}\\
	&\text{s.t.}\quad l^T_kv^*\leq c_0\alpha_c(r(q,l_k))-d_r,\quad k=1,\ldots,m\label{QPvirtualcontrolllaw2}\\
	&\qquad l^T_kv^*\leq c_0(\bar{v}-d_r)\quad k=1,\ldots,m\label{QPvirtualcontrolllaw3}
\end{align}
with a new class $\mathcal{K}^e$ function $\alpha_c$ and a distance-regularization function
\begin{align}
	r(q, l_k)=\min_{i=1,\ldots,N}\{|q_{oi}-q|-d_{si}+\phi( l_{oi}^T(q)l_k)\},\label{reshape}
\end{align}
where $\phi:[-1,1]\rightarrow \mathbb{R}_+$ is an appropriately chosen Lipschitz continuous, nonnegative,
decreasing function, $l_1,\ldots, l_m\in\mathbb{R}^n$ form a positive basis $\mathcal{L}$ such that any unit vector $l_o\in\mathbb{R}^n$ is a positive combination of the elements of $\{l_p: l_p^T l_o\geq c_0,~p=1,\ldots,m\}$ whose cardinality is not less than $n$ with constant $0<c_0<1$ and
\begin{align}
	l_{oi}(q)=\frac{q_{oi}-q}{|q_{oi}-q|}.\label{loidefine}
\end{align}

If the refined QP problem \eqref{QPvirtualcontrolllaw1}--\eqref{QPvirtualcontrolllaw3} admits a unique solution, then we use $\varphi:\mathbb{R}^n\times\mathbb{R}^n\rightarrow\mathbb{R}^n$ to represent the resulting map from $q$ and $v_c$ to $v^*$, i.e., 
\begin{align}
	v^*=\varphi(q,v_c).\label{varphidefine}
\end{align}

Denote
\begin{align}
	\underline{d}_s=\min_{i=1,\ldots,N}\{d_{si}\},~~~~\overline{d}_s=\max_{i=1,\ldots,N}\{d_{si}\}\label{dsupdown}.
\end{align}

\begin{figure}[h!]
	\centering
	\includegraphics[width=1\linewidth]{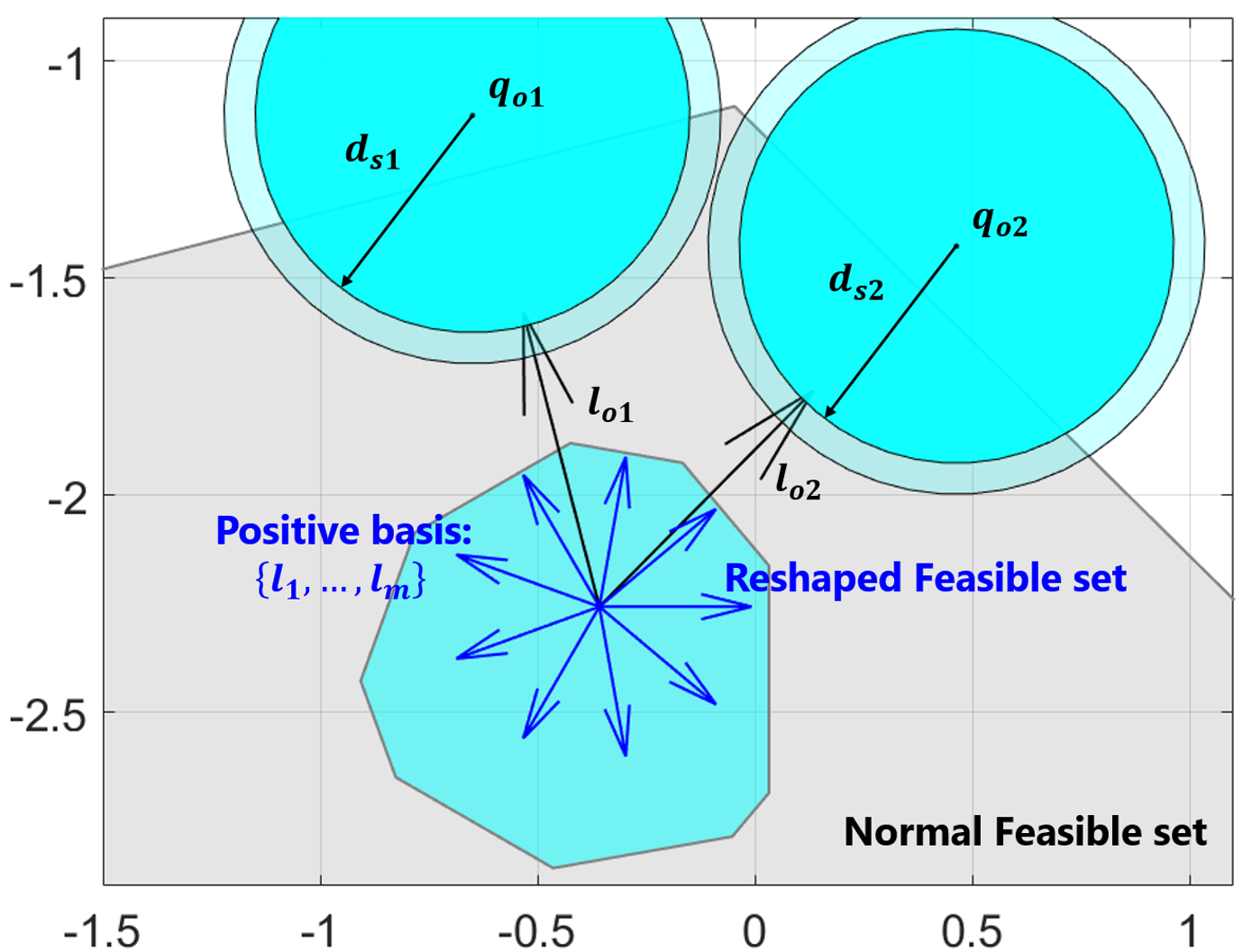}
	\caption{The basic idea of feasible-set reshaping.}
	\label{figure.feasiblesetreshape}
\end{figure}

Figure \ref{figure.feasiblesetreshape} illustrates the basic idea of the feasible-set reshaping technique. Proposition \ref{proposition.QP} shows that the QP-based control law with the proposed reshaped feasible set is capable of dealing with multiple position constraints while retaining the Lipschitz continuity.

\begin{proposition}\label{proposition.QP}
	Consider the kinematics part of the EL system \eqref{EL}:
	\begin{align}
		\dot{q}=v^*+\tilde{v}\label{integrator}
	\end{align}
	where $v^*\in\mathbb{R}^n$ is the solution of the QP problem \eqref{QPvirtualcontrolllaw1}--\eqref{QPvirtualcontrolllaw3} and $\tilde{v}\in\mathbb{R}^n$ represents the velocity-tracking error. Suppose that Assumptions \ref{assumption.vc} and \ref{assumption.balldistance} are satisfied. Under the conditions that
	\begin{align}
		\sqrt{2}\underline{d}_s&> d_a,\label{conditiona1}\\
		\bar{v}&>2d_r,\label{conditiona2}\\
		\frac{2\underline{d}_s^2-d_a^2}{2(\bar{d}_s+d_b/2)^2}&>1-\frac{(\bar{v}-2d_r)^2}{2(\bar{v}-d_r)^2}\label{conditiona3}
	\end{align}
	with $\bar{v}$ defined right after \eqref{barv}, $d_a$ and $d_b$ defined in Assumption \ref{assumption.balldistance}, and $d_r$ being a positive constant, there exist a Lipschitz continuous, nonnegative,
	decreasing function function $\phi:[-1,1]\rightarrow\mathbb{R}_+$, a constant $0<c_0<1$, and a Lipschitz function $\alpha_c\in\mathcal{K}^e$ whose inverse function is also Lipschitz satisfying
	\begin{align}
		&c_0>\sqrt{c^*}+\sqrt{c^*+\frac{d_r}{\bar{v}-d_r}},\label{conditionb1}\\
		&c_0(\bar{v}-d_r)\geq c_0\alpha_c\left(\frac{d_b}{2}\right)-d_r>\frac{d_r-c_0\alpha_c(-d_h)}{\sqrt{c^*+\frac{d_r}{\bar{v}-d_r}}-\sqrt{c^*}}, \label{conditionb2}\\
		&\phi(s)>\max_{p\in[0,d_h]}\Bigg\{\max_{K\in\bigg\{c_0(\bar{v}-d_r),\frac{d_r-c_0\alpha_c(-p)}{\sqrt{c^*+\frac{d_r}{\bar{v}-d_r}}-\sqrt{c^*}}	\bigg\}}\bigg\{\alpha_c^{-1}\bigg(\frac{-K\left(s-c_0\right)}{c_0}
		\notag\\
		&~~~~~~~~~~~~+\frac{-K\left(\sqrt{c^*+\frac{d_r}{\bar{v}-d_r}}-\sqrt{c^*}\right)+d_r}{c_0}\bigg)\bigg\}+p\Bigg\},~s\in[-1,c_0),\label{conditionb3}\\
		&\phi(s)=0,~ s\in[c_0,1]\label{conditionb4},
	\end{align}
	with
	\begin{align}
		c^*=\frac{(2\bar{d}_s+d_b)^2-4\underline{d}_s^2+2d_a^2}{2(2\bar{d}_s+d_b)^2},\label{cstar}
	\end{align}
	and $d_h$ defined in Assumption \ref{assumption.balldistance}. Moreover, there exists a constant $\Delta_h>0$ such that if
	\begin{align}
		h_i(q)\geq -\Delta_h,\quad i=1,\ldots,N,\label{QProbust}
	\end{align} 
	then
	\begin{enumerate}
		\item[(a)] the QP problem \eqref{QPvirtualcontrolllaw1}--\eqref{QPvirtualcontrolllaw3} admits a unique solution;
		\item[(b)] the position trajectories satisfy
		\begin{align}
			\frac{\partial h_i(q)}{\partial q}\dot{q}\geq -\alpha_p(h_i(q))+d_r-|\tilde{v}|,~ i=1,\ldots,N\label{gradienthi}
		\end{align}
		with
		\begin{align}
			\alpha_p(s)=\max\{\alpha_c(s),c_0\alpha_c(s)\},\label{alphap}
		\end{align}
		and the velocity reference signal $v^*$ satisfies
		\begin{align}
			|v^*|\leq\bar{v}-d_r;\label{barvP1}
		\end{align}
		\item[(c)] 
		the resulting function $\varphi$ in \eqref{varphidefine} generated by solving the QP problem \eqref{QPvirtualcontrolllaw1}--\eqref{QPvirtualcontrolllaw3} satisfies
		\begin{align}
			|\varphi(q_1,v_{c1})-\varphi(q_2,v_{c2})|\leq L(|q_1-q_2|+|v_{c1}-v_{c2}|)\label{L1L1P1}
		\end{align}
		for all $q_1,q_2$ satisfying \eqref{QProbust} and $v_{c1},v_{c2}\in\mathbb{R}^n$, where $L$ is a positive constant.
	\end{enumerate}
	
\end{proposition}

\begin{IEEEproof}
	Under conditions \eqref{conditiona1}--\eqref{conditiona3} and \eqref{cstar}, there exists a $c_0\in(0,1)$ satisfying \eqref{conditionb1}. Then, we have
	\begin{align}
		c_0(\bar{v}-d_r)>\frac{d_r}{\sqrt{c^*+d_r/(\bar{v}-d_r)}-\sqrt{c^*}},
	\end{align}
	which means the existence of an $\alpha_c\in\mathcal{K}^e$ satisfying \eqref{conditionb2}. Moreover, with $c_0$ determined, one can always find a $\phi$ satisfying \eqref{conditionb3}--\eqref{conditionb4}. 
	
	Now we prove the rest three properties of the proposition one-by-one.
	
	{\bfseries (a) Existence and Uniqueness of the Solution to the QP Problem}
	
	{\em (a.1) Existence of the Solution}
	
	Under Assumption \ref{assumption.balldistance}, with the conditions \eqref{conditiona1}--\eqref{cstar} satisfied, we choose $\Delta_h\in(0,d_h]$, $K_v>0$ and $c_M\in(0,1)$ satisfying
	\begin{align}
		&\frac{2(\underline{d}_s-\Delta_h)^2-d_a^2}{2(\bar{d}_s+d_b/2)^2}>c_M>1-\frac{(\bar{v}-2d_r)^2}{2(\bar{v}-d_r)^2},\label{conditionp2}\\
		&c_0>\sqrt{\frac{1-c_M}{2}}+\sqrt{\frac{1-c_M}{2}+\frac{d_r}{\bar{v}-d_r}},\label{conditionp3}\\
		&c_0(\bar{v}-d_r)\geq c_0\alpha_c\left(\frac{d_b}{2}\right)-d_r\geq K_v\geq\frac{-c_0\alpha_c(-\Delta_h)+d_r}{c_0-\sqrt{2-2c_M}},\label{conditionp4}\\
		&c_0-\sqrt{2-2c_M}>\sqrt{c^*+\frac{d_r}{\bar{v}-d_r}}-\sqrt{c^*},\label{conditionnew}\\
		&\phi(s)>\max_{K\in\left\{c_0(\bar{v}-d_r),\frac{d_r-c_0\alpha_c(-\Delta_h)}{\sqrt{c^*+\frac{d_r}{\bar{v}-d_r}}-\sqrt{c^*}}	\right\}}\Bigg\{\alpha_c^{-1}\Bigg(\frac{-K\left(s-c_0\right)}{c_0}
		\notag\\
		&~~~~~~~~~~~~+\frac{-K\left(\sqrt{c^*+\frac{d_r}{\bar{v}-d_r}}-\sqrt{c^*}\right)+d_r}{c_0}\Bigg)\Bigg\}+\Delta_h,~s\in[-1,c_0).\label{conditionp7}
	\end{align}
	In accordance with Assumption \ref{assumption.balldistance}, using condition \eqref{QProbust} and $\Delta_h\in(0,d_h]$, the definition of the distance-regularization function in \eqref{reshape} guarantees the existence of sets $K_1$ and $K_2$ satisfying
	\begin{align}
		K_1\cup K_2=\{1,\ldots,m\},~K_1\cap K_2=\emptyset
	\end{align}
	such that
	\begin{align}
		r(q,l_k)&\geq d_b/2,~for ~k\in K_1,\label{K1define}\\
		r(q,l_k)&<d_b/2,~for~k\in K_2\label{K2define}.
	\end{align}
	Then, we discuss the following two cases separately.
	\begin{itemize}
		\item $k\in K_1$. In this case, due to \eqref{conditionp4}, for any $v_s\in\mathbb{R}^n$ satisfying
		\begin{align}
			|v_s|\leq K_v,
		\end{align}
		we have
		\begin{align}
			l_k^Tv_s&\leq K_v\leq c_0\alpha_c(d_b/2)-d_r\notag\\
			&\leq c_0\alpha_c(r(q,l_k))-d_r
		\end{align}
		for all $k\in K_1$.
		
		\item $k\in K_2$. By using \eqref{conditionb3}--\eqref{conditionb4} and \eqref{conditionp2}--\eqref{conditionp7}, for $s\in[-1,1]$, we have
		\begin{align}
			\phi(s)\geq\max\left\{\alpha_c^{-1}\left(\frac{-K_v(s-\sqrt{2-2c_M})+d_r}{c_0}\right)+\Delta_h,0\right\}.\label{conditionp5}
		\end{align}
		In this case, from the definition of the distance-regularization function right in \eqref{reshape}, with the satisfaction of conditions \eqref{QProbust} and \eqref{conditionp5}, there exists a $J\subseteq\{1,\ldots,N\}$ such that for any $i\in J$,
		\begin{align}
			-\Delta_h\leq |q_{oi}-q|-d_{si}<d_b/2,
		\end{align}
		and for any $k\in K_2$, there exists an $i^*\in J$ such that
		\begin{align}
			r(q,l_k)=|q_{oi^*}-q|-d_{si^*}+\phi(l_{oi^*}^T(q)l_k)\geq-\Delta_h.\label{rkdelta}
		\end{align}
		With the satisfaction of Assumption \ref{assumption.balldistance}, under the condition that $\Delta_h\in(0,d_h]$, using \eqref{dsupdown}, \eqref{conditiona1}, \eqref{QProbust} and \eqref{conditionp2}, we have
		\begin{align}
			|q_{oi}-q_{oi'}|&\in [0,d_a],~\forall i,~i'\in J,\\
			l_{oi}^T(q) l_{oi'}(q)&=\frac{|q_{oi'}-q|^2+|q_{oi}-q|^2-|q_{oi}-q_{oi'}|^2}{2|q_{oi}-q||q_{oi'}-q|}\notag\\
			&\geq\frac{2(\underline{d}_s-\Delta_h)^2-d_a^2}{2(\bar{d}s+\frac{d_b}{2})^2}\geq c_M.\label{c0loiJ}
		\end{align}
		By using \eqref{c0loiJ}, we can prove that for any $i,i'\in J$,
		\begin{align}
			&|l_{oi}(q)-l_{oi'}(q)|\notag\\
			=&\sqrt{|l_{oi}(q)|^2+|l_{oi'}(q)|^2-2|l_{oi}(q)||l_{oi'}(q)|l_{oi}^T(q) l_{oi'}(q)}\notag\\
			\leq&\sqrt{2-2c_M}.\label{distanceloiJ}
		\end{align}
		Now, we show that $-K_v l_{oi}(q)$, for all $i\in J$, with $K_v$ defined in \eqref{conditionp4} satisfying the constraint \eqref{QPvirtualcontrolllaw2} for $k\in K_2$. With the satisfaction of conditions \eqref{conditionb4}, \eqref{conditionp5}, \eqref{rkdelta} and \eqref{distanceloiJ}, if $h_i(q)\geq -\Delta_h$ for $i=1,\ldots,N$, then
		\begin{align}
			-K_vl_{oi}^T(q)l_k
			&=-K_v l_k^T (l_{oi}(q)-l_{oi^*}(q)+l_{oi^*}(q))\notag\\
			&\leq  K_v\sqrt{2-2c_M}-K_vl_k^T l_{oi^*}(q)\notag\\
			&\leq c_0\alpha_c(\phi(l_k^T l_{oi^*}(q))-\Delta_h)-d_r\notag\\
			&\leq c_0\alpha_c(r(q, l_k))-d_r
		\end{align}
		holds for any $i\in J$ and $k\in K_2$. 
	\end{itemize}
	Thus, the QP problem \eqref{QPvirtualcontrolllaw1}--\eqref{QPvirtualcontrolllaw3} is always feasible.

	{\em (a.2) Uniqueness of the Solution}
	
	With the proved feasibility of the QP problem \eqref{QPvirtualcontrolllaw1}--\eqref{QPvirtualcontrolllaw3}, the uniqueness of the solution can be proved by directly applying the projection theorem; see \cite{Bertsekas-book-1997}.
	
	{\bfseries (b) Satisfaction of the Position and Velocity Constraints}
	
	Using the properties of positive basis and the definition of $\mathcal{L}$ right after \eqref{reshape}, for each $i=1,\ldots,N$, there exist $n$ elements of $\{ l_1,\ldots, l_m\}$, say $ l_{i1},\ldots, l_{in}$, and positive constants $\rho_{i1},\ldots,\rho_{in}$ such that
	\begin{align}
		l_{oi}(q)&=\sum_{j=1}^{n}\rho_{ij} l_{ij},\label{positiveexchange}\\
		l_{oi}^T(q) l_{ij}&\geq c_0,\quad j=1,\ldots,n.\label{loilijc0}
	\end{align}
	It can be directly verified that
	\begin{align}
		1\leq\sum_{j=1}^{n}\rho_{ij}\leq\frac{1}{c_0}.\label{maxmin}
	\end{align}
	With \eqref{positiveexchange} satisfied, we have
	\begin{align}
		l_{oi}^T(q)v^*=\sum_{j=1}^{n}\rho_{ij} l_{ij}^Tv^*,\quad i=1,\ldots,N.\label{loirholij}
	\end{align}
	Recall the definition of the distance-regularization function \eqref{reshape}. If conditions \eqref{QPvirtualcontrolllaw2}, \eqref{conditionb4} and \eqref{loilijc0} are satisfied, then for any element $l_{ij}$ of $\mathcal{L}$, we have
	\begin{align}
		l_{ij}^Tv^*&\leq c_0\alpha_c(r(q, l_{ij}))-d_r\notag\\
		&=c_0\alpha_c(\min_{w=1,\ldots,N}\{|q_{ow}-q|-d_{sw}+\phi( l_{ow}^T(q) l_{ij})\})-d_r\notag\\
		&\leq c_0\alpha_c(|q_{oi}-q|-d_{si})-d_r.\label{lijc0alpahac}
	\end{align}
	Then, \eqref{QPvirtualcontrolllaw2}, \eqref{maxmin}, \eqref{loirholij} and \eqref{lijc0alpahac} together imply
	\begin{align}
		l_{oi}^T(q)v^*&=\sum_{j=1}^{n}\rho_{ij} l_{ij}^Tv^*\notag\\
		&\leq\sum_{j=1}^{n}\rho_{ij}c_0\alpha_c(r(q, l_{ij}))-\sum_{j=1}^{n}\rho_{ij}d_r\notag\\
		&\leq\sum_{j=1}^{n}\rho_{ij}c_0\alpha_c(|q_{oi}-q|-d_{si})-\sum_{j=1}^{n}\rho_{ij}d_r\notag\\
		&\leq\sum_{j=1}^{n}\rho_{ij}c_0\alpha_c(|q_{oi}-q|-d_{si})-d_r.\label{safetyremakeo}
	\end{align}
	From \eqref{loilijc0} and \eqref{safetyremakeo}, we have that
	\begin{align}
		l_{oi}^T(q)v^*&\leq\begin{cases}
			\alpha_c(|q_{oi}-q|-d_{si})-d_r,~&\text{if}~|q_{oi}-q|-d_{si}\geq 0,\\
			c_0\alpha_c(|q_{oi}-q|-d_{si})-d_r,~&\text{if}~|q_{oi}-q|-d_{si}< 0.
		\end{cases}\label{safetyremakeo2}
	\end{align}
	Thus, we have
	\begin{align}
		l_{oi}^T(q)v^*\leq\alpha_p(|q_{oi}-q|-d_{si})-d_r\label{safetyremake}
	\end{align}
	where
	\begin{align}
		\alpha_p(s)=\max\{c_0\alpha_c(s),\alpha_c(s)\}
	\end{align}
	for $s\in\mathbb{R}$. It can be directly checked that $\alpha_p\in\mathcal{K}^e$.
	It is a direct consequence of \eqref{safetyremake} that \eqref{gradienthi} is satisfied.
	
	Suppose $v^*=k_x l_0$, with a positive constant $k_x$ and a unit vector $l_0\in\mathbb{R}^n$. Then, there exists at least one positive basis $l_i\in\mathcal{L}$ satisfying $ l_i^T l_0\geq c_0$ such that 
	\begin{align}
		k_x l_i^T l_0\leq c_0(\bar{v}-d_r),
	\end{align}
	which means
	\begin{align}
		|v^*|=k_x\leq\frac{c_0(\bar{v}-d_r)}{ l_i^T l_0}\leq\frac{c_0(\bar{v}-d_r)}{c_0}=\bar{v}-d_r.
	\end{align}

{\bfseries(c) Lipschitz Continuity and Boundedness of the QP-based Control Law}

Define
\begin{align}
	r_k(q,q_{oi},d_{si})=|q_{oi}-q|-d_{si}+\phi\left(l_k^T\frac{q_{oi}-q}{|q_{oi}-q|}\right).
\end{align}
Then, for any $q_1,q_2$ satisfying the constraints in \eqref{QProbust}, we have
\begin{align}
	&|r_k(q_1,q_{oi},d_{si})-r_k(q_2,q_{oi},d_{si})|\notag\\
	&=\left||q_{oi}-q_1|-|q_{oi}-q_2|+\phi(l_k^T\frac{q_{oi}-q_1}{|q_{oi}-q_1|})-\phi(l_k^T\frac{q_{oi}-q_2}{|q_{oi}-q_2|})\right|\notag\\
	&\leq |q_1-q_2|+L_{\phi}\left|\frac{q_{oi}-q_1}{|q_{oi}-q_1|}-\frac{q_{oi}-q_2}{|q_{oi}-q_2|}\right|\notag\\
	&\leq |q_1-q_2|+L_{\phi}\left|\frac{(q_{oi}-q_1)|q_{oi}-q_2|-(q_{oi}-q_2)|q_{oi}-q_1|}{|q_{oi}-q_1||q_{oi}-q_2|}\right|\notag\\
	&\leq |q_1-q_2| +L_{\phi}\Big|\frac{(q_{oi}-q_1-q_{oi}+q_2)|q_{oi}-q_2|}{|q_{oi}-q_1||q_{oi}-q_2|}\notag\\
	&~~~~-\frac{(q_{oi}-q_2)(|q_{oi}-q_1|-|q_{oi}-q_2|)}{|q_{oi}-q_1||q_{oi}-q_2|}\Big|\notag\\
	&\leq (1+\frac{2L_{\phi}}{d_{si}-\Delta_h})|q_1-q_2|
\end{align}
with $L_{\phi}>0$ being the Lipschitz constant of $\phi$. Intuitively, we have
\begin{align}
	r(q,l_k)=\min_{i=1,\ldots,N}\{r_k(q,q_{oi},d_{si})\}.
\end{align}
By using Proposition $2.3.9$ in \cite{Cobzas-Miculescu-Nicolae-book-2019}, we have
\begin{align}
	|r(q_2, l_k)-r(q_1, l_k)|&\leq \max_{i=1,\ldots,N}\left(1+\frac{2L_{\phi}}{d_{si}-\Delta_h}\right)|q_1-q_2|\notag\\
	&\leq L_r|q_1-q_2|,
\end{align}
where
\begin{align}
	L_r=1+\frac{2L_{\phi}}{\underline{d}_s-\Delta_h}.\label{Lr}
\end{align}

The constraints defined by \eqref{QPvirtualcontrolllaw2}--\eqref{QPvirtualcontrolllaw3} are equivalent to
\begin{align}
	l^T_iv^*\leq \min\{c_0\alpha_c(r(q, l_i))-d_r,c_0(\bar{v}-d_r)\},~k=1,\ldots,m,
\end{align}
where $\min\{c_0\alpha_c(r(q, l_i))-d_r,c_0(\bar{v}-d_r)\}$ is Lipschitz continuous with respect to $q$ for any $q$ satisfying $h_i(q)\geq 0$ for $i=1,\ldots,N$, and satisfies
\begin{align}
	&|\min\{c_0\alpha_c(r(q_1, l_i))-d_r,c_0(\bar{v}-d_r)\}\notag\\
	&-\min\{c_0\alpha_c(r(q_2, l_i))-d_r,c_0(\bar{v}-d_r)\}|\leq L_{\alpha}L_r|q_1-q_2|
\end{align}
where $L_{\alpha}$ is the Lipschitz constant of $c_0\alpha_c$.

Rewrite the left-hand side of \eqref{QPvirtualcontrolllaw2} as $A\varphi(q,v_c)$, where $A=[l_1^T\ldots l_m^T]$. Since $A$ is composed of unit vectors, $|A|$ is bounded. Since any $n$ rows of $A$ are linearly independent, there exists a positive constant $c_{\lambda}$ such that $\lambda_{\min}^{\frac{1}{2}}(AA^T)\geq c_{\lambda}$, and thus $|A^T\theta|\geq\lambda_{\min}^{\frac{1}{2}}(AA^T)|\theta|\geq c_{\lambda}|\theta|$ holds for all $\theta\in\mathbb{R}^n$; see \cite{Hager-SIAMControl-1979}.

With the guaranteed feasibility of the QP problem (\ref{QPvirtualcontrolllaw1})--(\ref{QPvirtualcontrolllaw3}), we can prove the Lipschitz continuity of the outer-loop control law following \cite{Hager-SIAMControl-1979}.

Moreover, we have
\begin{align}
	|\varphi(q_1,v_{c1})-\varphi(q_2,v_{c2})|\leq\rho|d(q_1,v_{c1})-d(q_2,v_{c2})|\label{QPLIPSCHITZ1}
\end{align}
where $d(q,v_c)=[-v_c^T,-\min\{c_0\alpha_c(r(q, l_1))-d_r,c_0(\bar{v}-d_r)\},\ldots,-\min\{c_0\alpha_c(r(q, l_m))-d_r,c_0(\bar{v}-d_r)\}]^T$ and $\rho=1/2+(1+4\sqrt{n}\max\{1/c_{\lambda},1\})/c_{\lambda}$, and thus
\begin{align}
	|\varphi(q_1,v_{c1})-\varphi(q_2,v_{c2})|\leq \rho |v_{c1}-v_{c2}|+\rho mL_{\alpha}L_{r}|q_1-q_2|.\label{QPLIPSCHITZ}
\end{align}
This proves the Lipschitz continuity of the solution $v^*$ of the QP problem with respect to $q$ satisfying $h_i(q)\geq -\Delta_h$ for $i=1,\ldots,N$ and $v_c$ satisfying $|v_c|\leq\bar{v}_c$.  Thus, there exists a positive constant $L$ such that \eqref{L1L1P1} is satisfied.
\end{IEEEproof}

\subsection{Velocity-Tracking Control Law}\label{section3b}

Given the velocity reference signal $v^*$ generated by the outer-loop control law proposed in Section \ref{section3a}, we design a velocity-tracking control law for the dynamics part of the system \eqref{EL}. Define
\begin{align}
\tilde{v}=\dot{q}-v^*\label{vstardotqtildev}
\end{align}
as the velocity-tracking error. Consider the following velocity-tracking control law 
\begin{align}
u=N(q,\dot{q})+C(q,\dot{q})v^*-k_D\tilde{v}\label{ic1}
\end{align}
with $k_D$ being a positive constant, and Lyapunov function candidate
\begin{align}
V(q,\tilde{v})=\frac{1}{2}\tilde{v}^TM(q)\tilde{v}.
\end{align}
Recall \eqref{varphidefine} and \eqref{vstardotqtildev}. We equivalently rewrite $V(q,\tilde{v})$ as
\begin{align}
\breve{V}(q,\dot{q},v_c)&=V(q,\tilde{v})\notag\\
&=\frac{1}{2}(\dot{q}-\varphi(q,v_c))^TM(q)(\dot{q}-\varphi(q,v_c)).\label{breveV}
\end{align}
The following proposition shows that the controlled EL system admits a velocity-tracking capability.
\begin{proposition}\label{innerloop}
Under Assumptions \ref{assumption.EL} and \ref{assumption.vc}, consider the system \eqref{EL} and the control law \eqref{ic1}. Suppose that there exists a positive constant $L$ such that the outer-loop control law represented by $v^*=\varphi(q,v_c)$ satisfies \eqref{L1L1P1} for all $q_1,v_{c1},q_2,v_{c2}\in\mathbb{R}^n$. Then, 
\begin{align}
	&\delta^T[\dot{q};\ddot{q};\dot{v}_c]\notag\\
	\leq&-\frac{2k_D}{\mu_{m2}}\breve{V}(q,\dot{q},v_c)+2nL\mu_{m2}\sqrt{\frac{2\breve{V}(q,\dot{q},v_c)}{\mu_{m1}}}(v_c^d+|\dot{q}|)\label{deltapartialP2}
\end{align}
holds for any $q,\dot{q},\ddot{q},v_c,\dot{v}_c\in\mathbb{R}^n$, and any $\delta\in\partial\breve{V}(q,\dot{q},v_c)$.
\end{proposition}
\begin{IEEEproof}
Rewrite the control law $\varphi(q,v_c)$ as
\begin{align}
	\varphi(q,v_c)=[\varphi_{c1}(q,v_c);\ldots;\varphi_{cn}(q,v_c)].\label{varphivrewrite}
\end{align}
With the existence of the positive constant $L$ for the satisfaction of \eqref{L1L1P1}, using Proposition 2.1.1 in \cite{Clarke-book-1990}, one can prove that
\begin{align}
	|\varphi_{ci}^{\circ}(q,v_{c},v)|\leq 2L|v|,~i=1,\ldots,n\label{P2p1}
\end{align}
for all $q,v_{c}\in\mathbb{R}^n$ and $v\in\mathbb{R}^{2n}$, where $\varphi_{ci}^{\circ}(q,v_{c},v)$ represents the directional derivative of $\varphi_{ci}(q,v_c)$. Then, according to the definition of generalized gradients in \eqref{genralizedgradient},
\begin{align}
	2L\geq|\epsilon|,~\forall \epsilon\in\partial \varphi_{ci}(q,v_c),~i\in\{1,\ldots,n\}\label{P2p2}
\end{align}
for $q,v_c\in\mathbb{R}^n$.
Denote
\begin{align}
	V_1(q,\dot{q},v_c)&=\dot{q}^TM(q)\dot{q},\label{P2V1}\\
	V_2(q,\dot{q},v_c)&=\dot{q}^TM(q)\varphi(q,v_c),\label{P2V2}\\
	V_3(q,\dot{q},v_c)&=\varphi(q,v_c)^TM(q)\varphi(q,v_c)\label{P2V3}.
\end{align}
Using the sum rule of generalized gradients (see \cite{Clarke-book-1990}) yields 
\begin{align}
	\partial\breve{V}(q,\dot{q},v_c)
	\subset\frac{1}{2}\partial V_1(q,\dot{q},v_c)-\partial V_2(q,\dot{q},v_c)+\frac{1}{2}\partial V_3(q,\dot{q},v_c).\label{P2Vr}
\end{align}
Under Assumptions \ref{assumption.EL} and \ref{assumption.vc}, with \eqref{sumset}, \eqref{EL}, \eqref{varphidefine}, \eqref{ic1}, \eqref{breveV} and \eqref{P2p2}--\eqref{P2Vr}, by using the product rule of generalized gradients in \cite{Clarke-book-1990}, we have
\begin{align}
	&\delta^T[\dot{q};\ddot{q};\dot{v}_c]\notag\\
	\leq&\dot{q}^TM(q)\ddot{q}+\frac{1}{2}\dot{q}^T\dot{M}(q)\dot{q}-\dot{q}^T\dot{M}(q)\varphi(q,v_c)-\varphi^T(q,v_c)M(q)\ddot{q}\notag\\
	&+2nL\mu_{m2}|\dot{q}-\varphi(q,v_c)|(|\dot{q}|+|\dot{v}_c|)+\frac{1}{2}\varphi^T(q,v_c)\dot{M}(q)\varphi(q,v_c)\notag\\
	\leq&-k_D|\dot{q}-\varphi(q,v_c)|^2+2nL\mu_{m2}|\dot{q}-\varphi(q,v_c)|(|\dot{q}|+|\dot{v}_c|)\notag\\
	\leq&-\frac{2k_D}{\mu_{m2}}\breve{V}(q,\dot{q},v_c)+2nL\mu_{m2}\sqrt{\frac{2\breve{V}(q,\dot{q},v_c)}{\mu_{m1}}}(v_c^d+|\dot{q}|),
\end{align}
for all $\delta\in\partial \breve{V}(q,\dot{q},v_c)$.
This proves the satisfaction of property \eqref{deltapartialP2} for any $q,\dot{q},v_c\in\mathbb{R}^n$, and any $\delta\in\partial\breve{V}(q,\dot{q},v_c)$.
\end{IEEEproof}

\subsection{Synthesis of the Inner-Outer-Loop Controller}\label{section3c}

In this section, we show the validity of the safety controller composed of the outer-loop safety control law in Section \ref{section3a} and the velocity-tracking control law in Section \ref{section3b}.

\begin{theorem}\label{theorem.P1P2}
Under Assumptions \ref{assumption.EL}, \ref{assumption.vc} and \ref{assumption.balldistance}, the closed-loop system composed of the EL system \eqref{EL} and inner-outer loop controller consisting of \eqref{QPvirtualcontrolllaw1}--\eqref{QPvirtualcontrolllaw3} and \eqref{ic1} admits the following properties: 
\begin{itemize}
	\item Under conditions \eqref{conditiona1}--\eqref{conditiona3}, with $\bar{v}$ defined right after \eqref{barv}, $d_a$ and $d_b$ defined in Assumption \ref{assumption.balldistance}, and $d_r$ being a positive constant, there exist a Lipshitz function $\phi:[-1,1]\rightarrow\mathbb{R}_+$, $0<c_0<1$ and Lipschitz $\alpha_c\in\mathcal{K}^e$ whose inverse function is also Lipschitz satisfying \eqref{conditionb1}--\eqref{conditionb4} with $c^*$ defined in \eqref{cstar} and $d_h$ defined in Assumption \ref{assumption.balldistance}, and the map $v^*=\varphi(q,v_c)$ representing the solution of \eqref{QPvirtualcontrolllaw1}--\eqref{QPvirtualcontrolllaw3} satisfies 
	\begin{align}
		&\min\{h_i(q_1),h_i(q_2)\}\geq-\Delta_h,~i=1,\ldots,N,~ q_1,q_2\in\mathbb{R}^n\notag\\
		\Rightarrow	&|\varphi(q_1,v_{c1})-\varphi(q_2,v_{c2})|\leq L(|q_1-q_2|+|v_{c1}-v_{c2}|)\label{T1lip}
	\end{align}
	for all $v_{c1},v_{c2}\in\mathbb{R}^n$, with positive constants $\Delta_h$ and $L$;
	\item In addition, with $k_D$ chosen such that
	\begin{align}
		k_D>2nL\frac{\mu_{m2}^3(v_c^d+\bar{v})}{\mu_{m1}^2d_r},\label{kD}
	\end{align}
	there exist a Lipschitz $\mathcal{K}^e$ function $\alpha_w$ and a positive constant $d_w$ such that 
	\begin{align}
		&W(q,\dot{q},v_c)\leq d_w\notag\\
		\Rightarrow&\delta^T[\dot{q};\ddot{q};\dot{v}_c]\leq -\alpha_w(W(q,\dot{q},v_v)),~\forall \delta\in\partial W(q,\dot{q},v_c),\label{T1aim}
	\end{align}
	holds for all
	$q,\dot{q},v_c\in\mathbb{R}^n$, where 	
	\begin{align}
		W(q,\dot{q},v_c)=\max\Big\{W_1(q,\dot{q},v_c),\ldots,W_{N+1}(q,\dot{q},v_c)\Big\},\label{maxW}
	\end{align}
	with
	\begin{align}
		W_i(q,\dot{q},v_c)&=-h_i(q),~i=1,\ldots,N,\label{Wi}\\
		W_{N+1}(q,\dot{q},v_c)&=-\frac{nL\mu_{m1}\mu_{m2}^2(v_c^d+\bar{v}-d_r)d_r}{\mu_{m1}k_D-2nL\mu_{m2}^2}\notag\\
		&~~~~+\frac{1}{2}(\dot{q}-\varphi(q,v_c))^TM(q)(\dot{q}-\varphi(q,v_c)).\label{WN1}
	\end{align}
\end{itemize}
\end{theorem}
\begin{IEEEproof}
Define
\begin{align}
	J_w(q,\dot{q},v_c)=\{i\in\{1,\ldots,N+1\}:W_i(q,\dot{q},v_c)=W(q,\dot{q},v_c)\}.\label{Jw}
\end{align}
Using the pointwise maxima rule and sum rule of generalized gradients in \cite{Clarke-book-1990}, we have
\begin{align}
	\partial W(q,\dot{q},v_c)\subset\conv\{\partial W_i(q,\dot{q},v_c),~i\in J_w(q,\dot{q},v_c)\}.\label{maxsub}
\end{align}
Under condition \eqref{maxsub}, using the definition of convex hull in \cite{Grunbaum-book-1967}, there exist appropriately chosen constants $\lambda_i$, for $i\in J_w(q,\dot{q},v_c)$  satisfying
\begin{align}
	\sum_{i\in J_w(q,\dot{q},v_c)}\lambda_i=1,~\min_{i\in J_w(q,\dot{q},v_c)}\lambda_i\geq 0,
\end{align}
and appropriately chosen vectors $\delta_i\in\partial W_i(q,\dot{q},v_c)$, $i\in J_w(q,\dot{q},v_c)$ such that for any $\delta\in\partial W(q,\dot{q},v_c)$,
\begin{align}
	\delta=\sum_{i\in J_w(q,\dot{q},v_c)}\lambda_i\delta_i.\label{deltasum}
\end{align}
The condition \eqref{kD} leads to
\begin{align}
	\frac{\mu_{m1}^2d_r^2}{2\mu_{m2}}-\frac{nL\mu_{m1}\mu_{m2}^2(v_c^d+\bar{v}-d_r)d_r}{\mu_{m1}k_D-2nL\mu_{m2}^2}>0.
\end{align}
Thus, we can choose $d_w$ satisfying
\begin{align}
	d_w\leq\min\left\{\Delta_h,\frac{\mu_{m1}^2d_r^2}{2\mu_{m2}}-\frac{nL\mu_{m1}\mu_{m2}^2(v_c^d+\bar{v}-d_r)d_r}{\mu_{m1}k_D-2nL\mu_{m2}^2}\right\}.\label{dw}
\end{align}
Under the condition that
\begin{align}
	W(q,\dot{q},v_c)\leq d_w,\label{Winf}
\end{align}
with Assumption \ref{assumption.EL} satisfied, using \eqref{maxW}, \eqref{WN1} and \eqref{Winf}, we have
\begin{align}
	|\dot{q}-\varphi(q,v_c)|&\leq \sqrt{\frac{\mu_{m1}}{\mu_{m2}}} d_r,\label{T1vtilde}\\
	h_i(q)&\geq -\Delta_h,~i=1,\ldots,N.\label{T1hi}
\end{align}
Consider the following two cases:
\begin{itemize}
	\item $N+1\notin J_w(q,\dot{q},v_c)$
	
In this case, with the satisfaction of Assumption \ref{assumption.EL}, from Proposition \ref{proposition.QP}, under conditions \eqref{Jw}--\eqref{T1hi} satisfied, there exists a Lipschitz, class $\mathcal{K}^e$ function $\alpha_{w1}$ such that
\begin{align}
	\delta^T[\dot{q};\ddot{q};\dot{v}_c]&=\sum_{i\in J_w(q,\dot{q},v_c)}-\lambda_i\nabla h_i(q)\dot{q}\notag\\
	&\leq\sum_{i\in J_w(q,\dot{q},v_c)}\lambda_i[\alpha_p(h_i(q))-d_r+|\dot{q}-\varphi(q,v_c)|]\notag\\
	&\leq\sum_{i\in J_w(q,\dot{q},v_c)}\lambda_i\alpha_p(h_i(q))\notag\\
	&=\sum_{i\in J_w(q,\dot{q},v_c)}-\lambda_i\alpha_{w1}(W_i(q,\dot{q},v_c))\notag\\
	&=-\alpha_{w1}(W(q,\dot{q},v_c))\label{T1dqh}
\end{align}
for any $\delta\in\partial W(q,\dot{q},v_c)$
\item $N+1\in J_w(q,\dot{q},v_c)$

By using techniques similar to those in \eqref{T1dqh}, we have for $\delta_i\in\partial W_i(q,\dot{q},v_c)$, $i\in J_w(q,\dot{q},v_c)/\{N+1\}$,
\begin{align}
	\delta_i^T[\dot{q};\ddot{q};\dot{v}_c]\leq -\alpha_{w1}(W(q,\dot{q},v_c)).
\end{align}
With Assumptions \ref{assumption.EL} and \ref{assumption.vc} satisfied, by using Propositions \ref{proposition.QP} and \ref{innerloop}, there exists a Lipschitz, class $\mathcal{K}^e$ function $\alpha_{w2}$ such that
\begin{align}
	&\delta_{N+1}^T[\dot{q};\ddot{q};\dot{v}_c]\notag\\
	\leq&-\frac{2k_D}{\mu_{m2}}\breve{V}(q,\dot{q},v_c)+2nL\mu_{m2}\sqrt{\frac{2\breve{V}(q,\dot{q},v_c)}{\mu_{m1}}}(v_c^d+|\dot{q}|)\notag\\
	\leq&-\frac{2k_D}{\mu_{m2}}\left(W_{N+1}(q,\dot{q},v_c)+\frac{nL\mu_{m1}\mu_{m2}^2(v_c^d+\bar{v}-d_r)d_r}{\mu_{m1}k_D-2nL\mu_{m2}^2}\right)\notag\\
	&~~+2nL\mu_{m2}\sqrt{\frac{2\breve{V}(q,\dot{q},v_c)}{\mu_{m1}}}\left(v_c^d+\bar{v}-d_r+\sqrt{\frac{2\breve{V}(q,\dot{q},v_c)}{\mu_{m1}}}\right)\notag\\
	\leq&-\frac{2k_D}{\mu_{m2}}\left(W_{N+1}(q,\dot{q},v_c)+\frac{nL\mu_{m1}\mu_{m2}^2(v_c^d+\bar{v}-d_r)d_r}{\mu_{m1}k_D-2nL\mu_{m2}^2}\right)\notag\\
	&~~+4nL\frac{\mu_{m2}}{\mu_{m1}}\left(W_{N+1}(q,\dot{q},v_c)+\frac{nL\mu_{m1}\mu_{m2}^2(v_c^d+\bar{v}-d_r)d_r}{\mu_{m1}k_D-2nL\mu_{m2}^2}\right)\notag\\
	&~~+2nL\mu_{m2}(v_c^d+\bar{v}-d_r)\sqrt{\frac{2\breve{V}(q,\dot{q},v_c)}{\mu_{m1}}}\notag\\
	=&-\left(\frac{2k_D}{\mu_{m2}}-4nL\frac{\mu_{m2}}{\mu_{m1}}\right)W_{N+1}(q,\dot{q},v_c)\notag\\
	&~~-\left(\frac{2k_D}{\mu_{m2}}-4nL\frac{\mu_{m2}}{\mu_{m1}}\right)\frac{nL\mu_{m1}\mu_{m2}^2(v_c^d+\bar{v}-d_r)d_r}{\mu_{m1}k_D-2nL\mu_{m2}^2}\notag\\
	&~~+2nL\mu_{m2}(v_c^d+\bar{v}-d_r)\sqrt{\frac{2\breve{V}(q,\dot{q},v_c)}{\mu_{m1}}}\notag\\
	\leq&-\left(\frac{2k_D}{\mu_{m2}}-4nL\frac{\mu_{m2}}{\mu_{m1}}\right)W_{N+1}(q,\dot{q},v_c)\notag\\
	&~~+2nL\mu_{m2}(v_c^d+\bar{v}-d_r)\left(\sqrt{\frac{2\breve{V}(q,\dot{q},v_c)}{\mu_{m1}}}-d_r\right)\notag\\
	\leq&-\left(\frac{2k_D}{\mu_{m2}}-4nL\frac{\mu_{m2}}{\mu_{m1}}\right)W_{N+1}(q,\dot{q},v_c)\notag\\
	\leq&-\alpha_{w2}(W(q,\dot{q},v_c))
\end{align}
for any $\delta_{N+1}\in\partial W_{N+1}(q,\dot{q},v_c)$, where we have used conditions \eqref{vstardotqtildev}, \eqref{breveV}, \eqref{T1lip}, \eqref{maxW}--\eqref{WN1}, and \eqref{Winf}--\eqref{T1hi} for the first inequality, used conditions \eqref{matrix}, \eqref{barvP1} \eqref{maxW}--\eqref{WN1}, and \eqref{Winf}--\eqref{T1hi} for the second inequality, used condition \eqref{WN1} for the third inequality, and used conditions \eqref{matrix}, \eqref{kD}, \eqref{maxW} and \eqref{Winf}--\eqref{T1vtilde} for the fifth inequality.

Thus, for any $\delta_i\in W_i(q,\dot{q},v_c)$, $i\in J_w(q,\dot{q},v_c)$, we have
\begin{align}
	\delta_i^T[\dot{q};\ddot{q};\dot{v}_c]\leq-\min\left\{\alpha_{w1}(W(q,\dot{q},v_c)),\alpha_{w2}(W(q,\dot{q},v_c))\right\}.\label{deltaicase2}
\end{align}
From conditions \eqref{Jw}--\eqref{T1hi} and \eqref{deltaicase2}, we have for any $\delta\in \partial W(q,\dot{q},v_c)$,
\begin{align}
	\delta^T[\dot{q};\ddot{q};\dot{v}_c]&=\sum_{i\in J_w(q,\dot{q},v_c)}\lambda_i\delta_i^T[\dot{q};\ddot{q};\dot{v}_c]\notag\\
	&\leq-\min\left\{\alpha_{w1}(W(q,\dot{q},v_c)),\alpha_{w2}(W(q,\dot{q},v_c))\right\}.\label{T1gradientp1}
\end{align}
\end{itemize}
Thus, \eqref{T1aim} is obtained by choosing
\begin{align}
\alpha_w(s)=\min\left\{\alpha_{w1}(s),\alpha_{w2}(s)\right\}.\label{T1gradientp2}
\end{align}

\end{IEEEproof}
%

\begin{remark}\label{remark.safety}
For a max-type Lyapunov-like function \eqref{maxW}, with condition \eqref{T1aim} satisfied, using Lemma $2.15$ in \cite{Della-2020-nonsmooth}, we have
\begin{align}
\dot{W}(q(t),\dot{q}(t),v_c(t)) \leq -\alpha_w(W(q(t),\dot{q}(t),v_c(t)))\label{T1ae}
\end{align}
for almost every $t\geq 0$ such that
\begin{align}
W(q(t),\dot{q}(t),v_c(t))\leq d_w.\label{T1aen}
\end{align}
With Theorem $1.10.2$ in \cite{Lakshmikantham-book-1969}, there exists a $\beta\in\mathcal{KL}$ satisfying $\beta(s,0)=s$ for all $s\geq 0$ such that under the initial condition $ W(q(0),\dot{q}(0),v_c(0))\leq d_w$,
\begin{align}
&W(q(t),\dot{q}(t),v_c(t))\notag\\
\leq& \sgn(W(q(0),\dot{q}(0),v_c(0)))\beta(|W(q(0),\dot{q}(0),v_c(0))|,t),~\forall t\geq 0.\label{Winv}
\end{align}
Under conditions \eqref{conditiona1}--\eqref{conditiona3}, \eqref{kD}, \eqref{maxW}--\eqref{WN1} and \eqref{Winv}, according to Assumptions \ref{assumption.EL}--\ref{assumption.balldistance}, using Proposition \ref{proposition.QP}, we have that if
\begin{align}
W(q(0),\dot{q}(0),v_c(0))\leq 0,\label{remarkinitialv}
\end{align}
then \eqref{safetyregion} and \eqref{barv} are satisfied for all $t\geq 0$.
\end{remark}

\begin{remark}
The max-type combination of the barrier functions and the Lyapunov function of different subsystems of the closed-loop system is somehow motivated by the Lyapunov-based nonlinear small-gain theorem (\cite{Jiang-Mareels-Wang-Auto-1996}), in which max-type Lyapunov functions are constructed for interconnected systems based on Lyapunov functions of the subsystems.
\end{remark}

\section{Refined Control Law with a Possibly Reduced Neighborhood}\label{T2}

The QP-based control law in the previous section takes into account all obstacles, the number of which could be very large. The computation load required by safety control would be significantly increased as the number of obstacles becomes large. This section shows that the obstacles with positions far away enough from the plant state do not influence the position constraints much, and proposes a refined control law to reduce the computation load.

Recall the function $\phi$ and the unit vectors $l_k$ for $k=1,\ldots,m,$ defined right after \eqref{reshape}, and $l_{oi}(q)$ defined in \eqref{loidefine}. We replace the distance-regularization function $r$ in \eqref{reshape} with
\begin{align}
r_f(q,l_k,d_f)&=\min\left\{\min_{i\in J_N(q,d_f)}\{|q_{oi}-q|-d_{si}+\phi(l_{oi}^T(q)l_k)\},d_f\right\},\label{newreshape}
\end{align}
where
\begin{align}
J_N(q,d_f)&=\{i\in\{1,\ldots,N\}:|q_{oi}-q|-d_{si}\leq d_f\}\label{JN}
\end{align}
with $d_f$ being a positive constant to be determined later. Then, for $k=1,\ldots,m$, we define
\begin{align}
r_k(q,d_f)&=\begin{cases}
r_f(q,l_k,d_f),~&\text{if}~J_N(q,d_f)\neq\emptyset,\\
d_f,~&\text{if}~J_N(q,d_f)=\emptyset,
\end{cases}\label{rkqdf}
\end{align}
and consider the following refined QP-based control law:
\begin{align}
& v^*=\mathop{\argmin}\limits_{v^*\in\mathbb{R}^n}|v^*-v_c|^2\label{QPvirtualcontrolllaw1n}\\
&\text{s.t.}\quad l^T_kv^*\leq c_0\alpha_c(r_k(q,d_f))-d_r,\quad k=1,\ldots,m\label{QPvirtualcontrolllaw2n}
\end{align}
where $\alpha_c$ is a class $\mathcal{K}^e$ function.

The following proposition shows that the refined QP-based control law which only considers the obstacles nearby is still capable of dealing with all the position constraints and guaranteeing the required Lipschitz continuity.
\begin{proposition}\label{proposition.QPnearby}
Consider the kinematics part of the EL system \eqref{EL}:
\begin{align}
\dot{q}=v^*+\tilde{v}\label{integrator}
\end{align}
where $v^*\in\mathbb{R}^n$ is the solution of the QP problem \eqref{QPvirtualcontrolllaw1n}--\eqref{QPvirtualcontrolllaw2n} and $\tilde{v}\in\mathbb{R}^n$ represents the generalized velocity-tracking error. Suppose that Assumptions \ref{assumption.vc} and \ref{assumption.balldistance} are satisfied. Under conditions \eqref{conditiona1}--\eqref{conditiona3}, where $\bar{v}$ is defined right after \eqref{barv}, $d_a$ and $d_b$ are defined in Assumption \ref{assumption.balldistance}, and $d_r$ is a positive constant, there exist a Lipschitz continuous, nonnegative,
decreasing function $\phi:[-1,1]\rightarrow\mathbb{R}_+$, a positive constant $0<c_0<1$, a positive constant $d_f$ and a Lipschitz function $\alpha_c\in\mathcal{K}^e$ satisfying \eqref{conditionb1}--\eqref{conditionb4} and
\begin{align}
&\alpha_c^{-1}\left(\frac{-c_0\alpha_c(-d_h)+\left(1+\sqrt{c^*+\frac{d_r}{\bar{v}-d_r}}-\sqrt{c^*}\right)d_r}{c_0\left(\sqrt{c^*+\frac{d_r}{\bar{v}-d_r}}-\sqrt{c^*}\right)}\right)\notag\\
&<d_f\leq \alpha_c^{-1}\left(\bar{v}+(\frac{1}{c_0}-1)d_r\right),\label{df}
\end{align}
with $c^*$ defined in \eqref{cstar} and $d_h$ defined in Assumption \ref{assumption.balldistance}. Moreover, there exists a constant $\Delta_h>0$ such that if
\begin{align}
h_i(q)\geq -\Delta_h,\quad i=1,\ldots,N,\label{QProbust1}
\end{align} 
then
\begin{enumerate}
\item[(a)] the QP problem \eqref{QPvirtualcontrolllaw1n}--\eqref{QPvirtualcontrolllaw2n} admits a unique solution;
\item[(b)] the position trajectories satisfy
\begin{align}
	\frac{\partial h_i(q)}{\partial q}\dot{q}\geq -\alpha_p(h_i(q))+d_r-|\tilde{v}|,~ i=1,\ldots,N\label{gradienthi1}
\end{align}
with
\begin{align}
	\alpha_p(s)=\max\{\alpha_c(s),c_0\alpha_c(s)\},\label{alphap1}
\end{align}
and the velocity reference signal $v^*$ satisfies
\begin{align}
	|v^*|\leq\bar{v}-d_r;
\end{align}
\item[(c)] the map $\varphi$ in \eqref{varphidefine} generated by solving the QP problem \eqref{QPvirtualcontrolllaw1n}--\eqref{QPvirtualcontrolllaw2n} satisfies
\begin{align}
	|\varphi(q_1,v_{c1})-\varphi(q_2,v_{c2})|\leq L(|q_1-q_2|+|v_{c1}-v_{c2}|)\label{L1L1P11}
\end{align}
for all $q_1,q_2$ satisfying \eqref{QProbust1} and $v_{c1},v_{c2}\in\mathbb{R}^n$, where $L$ is a positive constant.
\end{enumerate}

\end{proposition}
\begin{IEEEproof}[Sketch of Proof]
Conditions \eqref{conditiona3}--\eqref{conditionb2} readily guarantees the existence of $d_f$ satisfying \eqref{df}.

{\bfseries (a) Existence and Uniqueness of the Solution to the QP Problem}

{\em (a.1) Existence of the Solution}

Conditions \eqref{conditiona1}--\eqref{cstar} and \eqref{df} guarantee that there exist constants $K_v>0$, $c_M\in(0,1)$ and $\Delta_h\in(0,d_h]$ satisfying
\begin{align}
c_0(\bar{v}-d_r)&\geq c_0\alpha_c\left(\max\left\{d_f,\frac{d_b}{2}\right\}\right)-d_r\notag\\
& \geq c_0\alpha_c\left(\min\left\{d_f,\frac{d_b}{2}\right\}\right)-d_r\geq K_v\geq \frac{-c_0\alpha_c(-\Delta_h)+d_r}{c_0-\sqrt{2-2c_M}}, \label{conditionKv1}\\
\phi(s)&\geq\alpha_c^{-1}\left(\frac{-K_v(s-\sqrt{2-2c_M})+d_r}{c_0}\right)+\Delta_h.\label{phiP4}
\end{align}
According to conditions \eqref{rkqdf}, \eqref{conditionKv1}--\eqref{phiP4}, using the proof similar to Proposition \ref{proposition.QP}, there exists a $v_x\in\mathbb{R}^n$ satisfying
\begin{align}
|v_x|= K_v,\label{vxP4}
\end{align}
such that for all $k=1,\ldots,m$, if $J_N(q,d_f)\neq\emptyset$, then
\begin{align}
l^T_kv_x&\leq c_0\alpha_c(r_f(q,l_k,d_f))-d_r,\label{dfconstraint}
\end{align}
and if $J_N(q,d_f)=\emptyset$, then
\begin{align}
l_k^Tv_x\leq c_0\alpha_c(d_f)-d_r\leq c_0(\bar{v}-d_r).\label{Kvdf}
\end{align}
Thus, the QP problem \eqref{QPvirtualcontrolllaw1n}--\eqref{QPvirtualcontrolllaw2n} is always feasible.

{\em (a.2) Uniqueness of the Solution}

With the proved feasibility of the QP problem \eqref{QPvirtualcontrolllaw1n}--\eqref{QPvirtualcontrolllaw2n}, the uniqueness for the solution can be proved by directly applying the projection theorem; see \cite{Bertsekas-book-1997}.

{\bfseries (b) Satisfaction of the Position and Velocity Constraints}

Suppose that $v^*=k_x l_0$, where $k_x$ is a positive constant and $l_0\in\mathbb{R}^n$ is some unit vector. Then, due to \eqref{JN}--\eqref{QPvirtualcontrolllaw2n} and the definition of positive basis right after \eqref{reshape}, there exists at least one positive basis $l_k\in\mathcal{L}$ satisfying $ l_k^T l_0\geq c_0$ such that 
\begin{align}
	k_x l_k^T l_0\leq c_0\alpha_c(d_f)-d_r.\label{kxlilo}
\end{align}
From \eqref{conditionKv1}, \eqref{Kvdf} and \eqref{kxlilo}, we have
\begin{align}
	|v^*|=k_x\leq\frac{c_0\alpha_c(d_f)-d_r}{ l_k^T l_0}\leq\alpha_c(d_f)-\frac{d_r}{c_0}=\bar{v}-d_r.\label{bound1}
\end{align}
Obviously, the solution to the QP problem is bounded according to \eqref{bound1}. Moreover, for any $i\in\{1,\ldots,N\}/J_N(q,d_f)$, by using \eqref{JN}, \eqref{alphap1}, \eqref{conditionKv1}, \eqref{bound1} and $0<c_0<1$, we have
\begin{align}
	l_{oi}^T(q)v^*&\leq\alpha_c(d_f)-\frac{d_r}{c_0}\notag\\
	&\leq\alpha_c(|q_{oi}-q|-d_{si})-d_r\notag\\
	&=\alpha_p(|q_{oi}-q|-d_{si})-d_r\label{safetynb1},
\end{align}
which means that 
\begin{align}
	\frac{\partial h_i(q)}{\partial q}\dot{q}\geq -\alpha_p(h_i(q))+d_r-|\tilde{v}|,~ \forall i\in\{1,\ldots,N\}/J_N(q,d_f).
\end{align}

Applying the same techniques as for the proof of Proposition \ref{proposition.QP}, we have
\begin{align}
	l_{oi}^T(q)v^*\leq\alpha_p(|q_{oi}-q|-d_{si})-d_r,~i\in J_N(q,d_f).\label{safetynb2}
\end{align}
It is a direct consequence of \eqref{safetynb1} and \eqref{safetynb2} that \eqref{gradienthi1} is satisfied.

{\bfseries(c) Lipschitz Continuity and Boundedness of the Refined QP-based Control Law}

From \eqref{reshape}, \eqref{JN} and \eqref{newreshape}, for any $q\in\mathbb{R}^n$, if $r_f(q,d_f,l_k)=d_f$, we have
\begin{align}
	r(q,l_k)\geq d_f,
\end{align}
and if $r_f(q,d_f,l_k)=\min_{i\in J_N(q,d_f)}\{|q_{oi}-q|-d_{si}+\phi(l_{oi}^T(q)l_k)\}$, we have
\begin{align}
	r_f(q,d_f,l_k)=r(q,l_k)<d_f.
\end{align}
Thus, for any $q_1,q_2\in\mathbb{R}^n$ satisfying \eqref{QProbust1}, constant $d_f$ satisfying \eqref{df} and $l_k$ for $k=1,\ldots,m$ defined right after \eqref{reshape},
\begin{align}
	|r_f(q_1,d_f)-r_f(q_2,d_f)|\leq|r(q_1,l_k)-r(q_2,l_k)|.
\end{align}
Then, we can prove the existence of a positive constant $L$ for \eqref{L1L1P11} using the same techniques as for the proof of Proposition \ref{proposition.QP}.
\end{IEEEproof}

The following theorem proposes a controller for constraint satisfaction based on Proposition \ref{proposition.QPnearby} and Proposition \ref{innerloop}.

\begin{theorem}\label{theorem.P2P3}
Under Assumptions \ref{assumption.EL}, \ref{assumption.vc} and \ref{assumption.balldistance}, the closed-loop system composed of the EL system \eqref{EL} and inner-outer loop controller consisting of \eqref{QPvirtualcontrolllaw1n}--\eqref{QPvirtualcontrolllaw2n} and \eqref{ic1} admits the following properties: 
\begin{itemize}
	\item Under conditions \eqref{conditiona1}--\eqref{conditiona3}, with $\bar{v}$ defined right after \eqref{barv}, $d_a$ and $d_b$ defined in Assumption \ref{assumption.balldistance}, and $d_r$ being a positive constant, there exist a Lipshitz function $\phi:[-1,1]\rightarrow\mathbb{R}_+$, $0<c_0<1$ and Lipschitz $\alpha_c\in\mathcal{K}^e$ whose inverse function is also Lipschitz satisfying \eqref{conditionb1}--\eqref{conditionb4} and \eqref{df} with $c^*$ defined in \eqref{cstar} and $d_h$ defined in Assumption \ref{assumption.balldistance}, and the map $v^*=\varphi(q,v_c)$ representing the solution of \eqref{QPvirtualcontrolllaw1n}--\eqref{QPvirtualcontrolllaw2n} satisfies \eqref{T1lip} for all $v_{c1},v_{c2}\in\mathbb{R}^n$, with positive constants $\Delta_h$ and $L$;
	\item In addition, with $k_D$ satisfying \eqref{kD}, there exists a Lipschitz $\mathcal{K}^e$ function $\alpha_w$ and a positive constant $d_w$ such that \eqref{T1aim}--\eqref{WN1} holds for all $q,\dot{q},v_c\in\mathbb{R}^n$.
\end{itemize}
\end{theorem}

The discussion on safety in Remark \ref{remark.safety} still applies to this case.

\section{Numerical and Experimental Validations}
\label{section.simulation}

In typical scenarios of robotic manipulator control, it is essential that every point on the body keeps far away enough from unsafe regions. With $\mathcal{B}\subset\mathbb{R}^n$ representing the unsafe regions in the workspace defined in $\mathbb{R}^n$ and $g:\mathbb{R}^n\leadsto\mathbb{R}^n$ mapping the unsafe region from the workspace to the configuration space defined in $\mathbb{R}^n$, we can solve the safety-critical control problem by directly dealing with the unsafe regions $g(\mathcal{B})$ in the configuration space.

For manipulators working in cluttered environments, the unsafe regions may not be explicitly represented by known functions. Here, we use a set of $N$ balls to cover the unsafe regions in the configuration space, and solve the safety-critical control problem by considering the balls as obstacles:
\begin{align}
\bigcup_{i=1}^N\{q\in\mathbb{R}^n:|q-q_{oi}|-d_{si}\leq 0\}\label{rasterize}
\end{align}
where $q_{oi}\in\mathbb{R}^n$ and $d_{si}>0$ are the center and the radius of the $i$-th ball, respectively.

\begin{figure}[h!]
\centering
\includegraphics[width=1\linewidth]{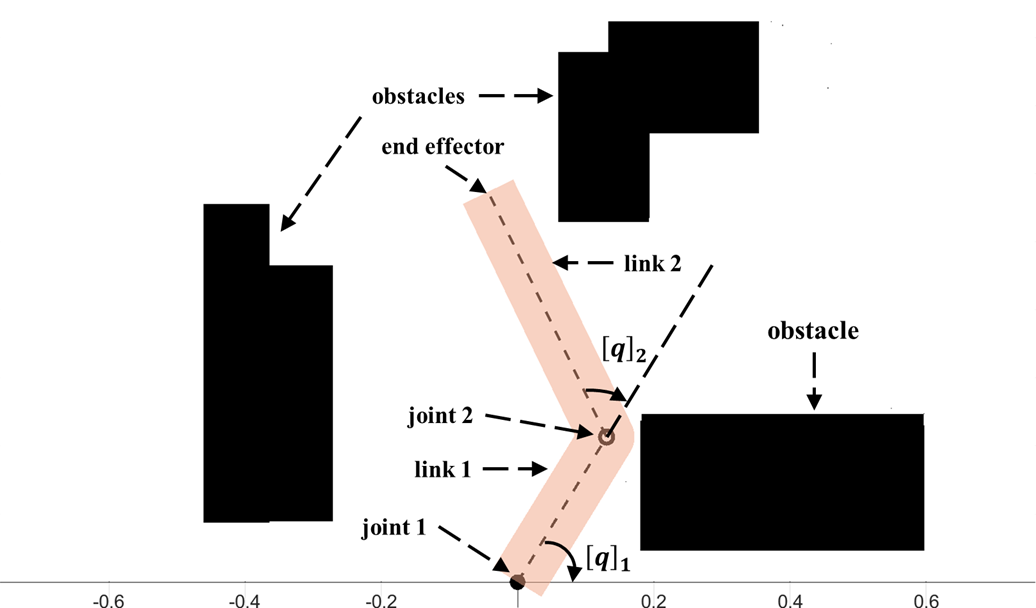}
\caption{A scenario of safety-critical control of a $2$-link robotic manipulator.}
\label{figure.simulationscenario}
\end{figure}

We consider a scenario of safety-critical control of a $2$-link planar manipulator, as shown in Figure \ref{figure.simulationscenario}. Safety essentially requires that every point on the body of the $2$-link planar manipulator keeps far away enough from the obstacles. Moreover, with $P_2\in\mathbb{R}^2$ and $P_e\in\mathbb{R}^2$ being the positions of joint $2$ and end effector, safety also requires that $[P_2]_2\geq 0$ and $[P_e]_2\geq 0$. Link $1$ is a uniform rod with width $0.07$ m, length $0.25$ m and mass $2$ kg; link $2$ is a uniform rod with width $0.07$ m, length $0.4$ m and mass $3$ kg. SI units are used in this section, and the units are omitted where it does not lead to confusion.

\subsection{A Safety-Critical Control Scenario and Controller Design}\label{controllerse}

The positions $q_{oi}$ of the ball obstacles in the joint (configuration) space are generated by an algorithm based on the following idea: The joint space is first divided into uniform-sized hypercubes, and balls with
the same radius are assigned to the hypercubes that intersect with the unsafe region. In particular, the side length of the hypercubes is chosen as $0.0698$ rad and the radius of the balls as $0.1396$ rad.

We design two controllers whose outer-loop safety control laws are in the form of \eqref{QPvirtualcontrolllaw1}--\eqref{QPvirtualcontrolllaw3} and \eqref{QPvirtualcontrolllaw1n}--\eqref{QPvirtualcontrolllaw2n}, respectively. For both of the controllers, we choose
\begin{align*}
d_a&=0.1561,\hspace{10pt} d_b=0.0182,\hspace{10pt} d_r=0.0105,\hspace{10pt}
d_h=0.00002,\notag\\
d_{si}&=0.1745,\hspace{10pt} c_0=0.9965,\hspace{10pt} \bar{v}~~=0.3142,\notag\\
\alpha_c(r)&=\begin{cases}
	0.9r,~&\text{for}~r\leq 0,\\
	34.5506r,~&\text{for}~0<r\leq 0.0091,\\
	0.9(r-0.0091)+0.3136,~&\text{for}~0.0091<r,
\end{cases}\\
\phi(r)&=\max\{-0.3362s+0.3350,0\},~r\in[-1,1].
\end{align*}
Moreover, for the controller whose outer-loop safety control law is in the form of \eqref{QPvirtualcontrolllaw1n}--\eqref{QPvirtualcontrolllaw2n}, we choose
\begin{align}
	d_f=0.0098.\notag
\end{align}
Then, Assumption \ref{assumption.balldistance} is verified. 

\subsection{Numerical Simulation}

The coefficients of the EL model \eqref{EL} for the robotic manipulator can be readily calculated following the standard modeling approach \cite{Craig-book-2005}. Regarding the $2$-link planar manipulator, the coefficients are available in \cite{Mahil-Ahmed-MWSCAS-2016}. Moreover, we design the inner-loop velocity-tracking control law in the form of \eqref{ic1} with $k_D=800.0000$.

The primary controller is designed based on saturated proportional feedback and an appropriately chosen waypoint to keep the primary velocity command within the desired range and avoid deadlock:
\begin{align}
	v_c=-(q-q_{ri})\min\left\{0.5236,\frac{\bar{v}_c}{|q-q_{ri}|}\right\}
\end{align}
with $\bar{v}_c$ being the upper bound of the velocity command signal $v_c$, $q_{r1}$ being the waypoint, and $q_{r2}$ being the target state. In this numerical simulation, the state $q$ approaches from the initial state $q(0)=[1.5208,-1]^T$ to the target state $q_{r2}=[1.7208,0]^T$ through the waypoint $q_{r1}=[2.6,-1.6]^T$. Here, we set $\bar{v}_c=0.2618$.

\begin{figure}[h!]
	\centering
	\includegraphics[width=1\linewidth]{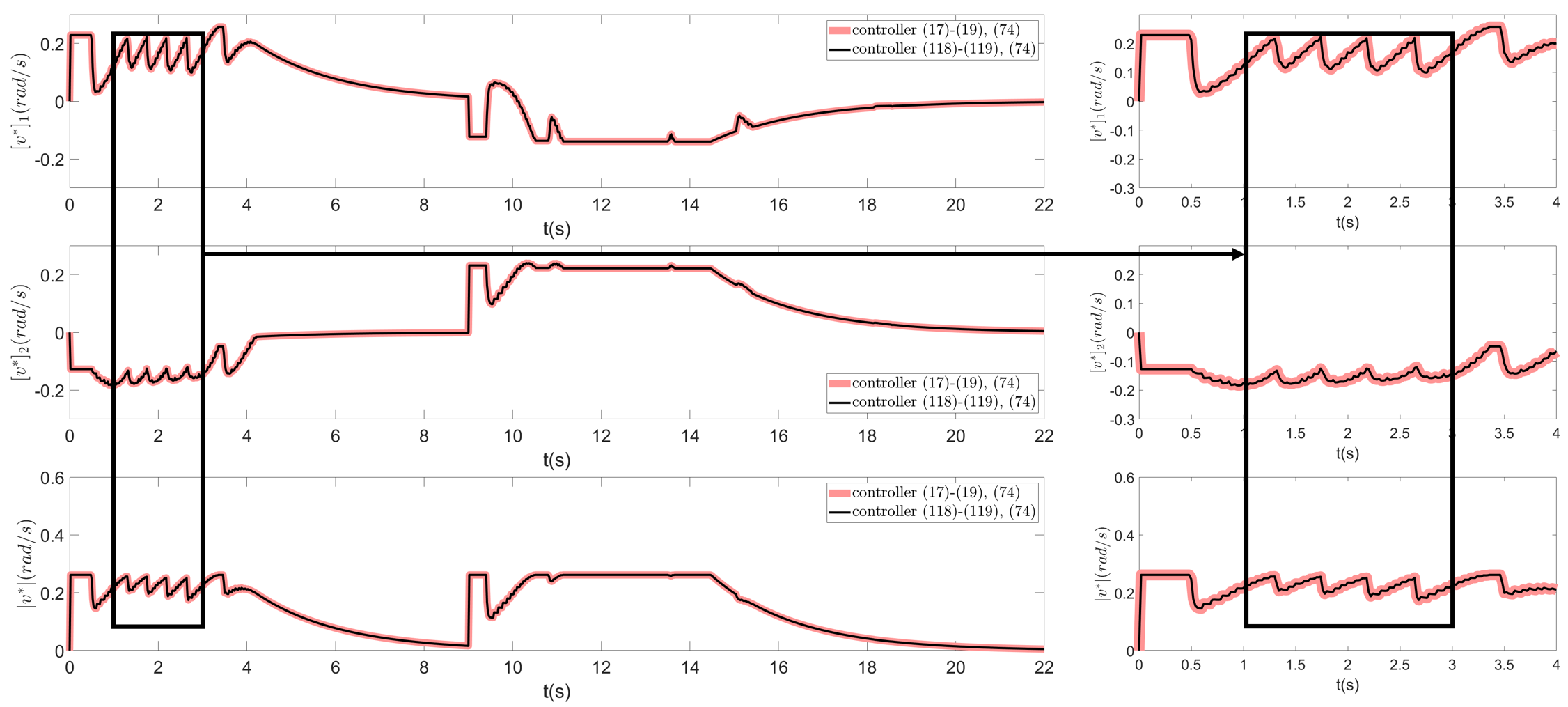}
	\caption{Trajectories of the velocity reference signal $v^*$ in the numerical simulation.}
	\label{figure.vstartsim}
\end{figure}

\begin{figure}[h!]
	\centering
	\includegraphics[width=1\linewidth]{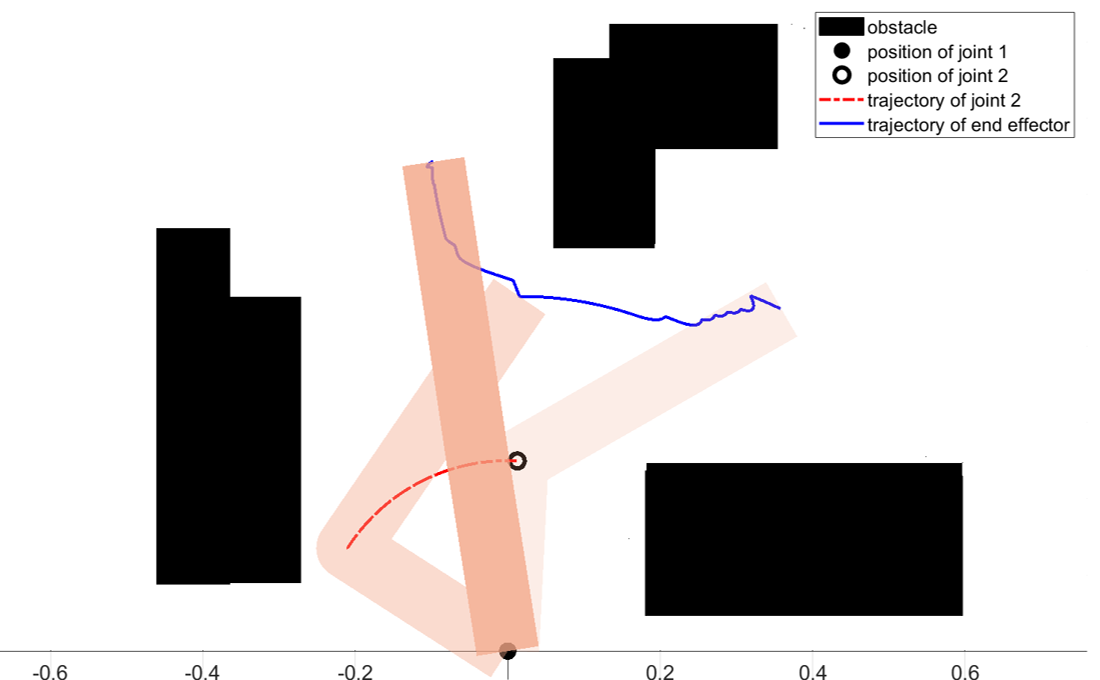}
	\caption{The motion of the $2$-link robotic manipulator in the workspace.}
	\label{figure.simulationworkspace}
\end{figure}

\begin{figure}[h!]
	\centering
	\includegraphics[width=1\linewidth]{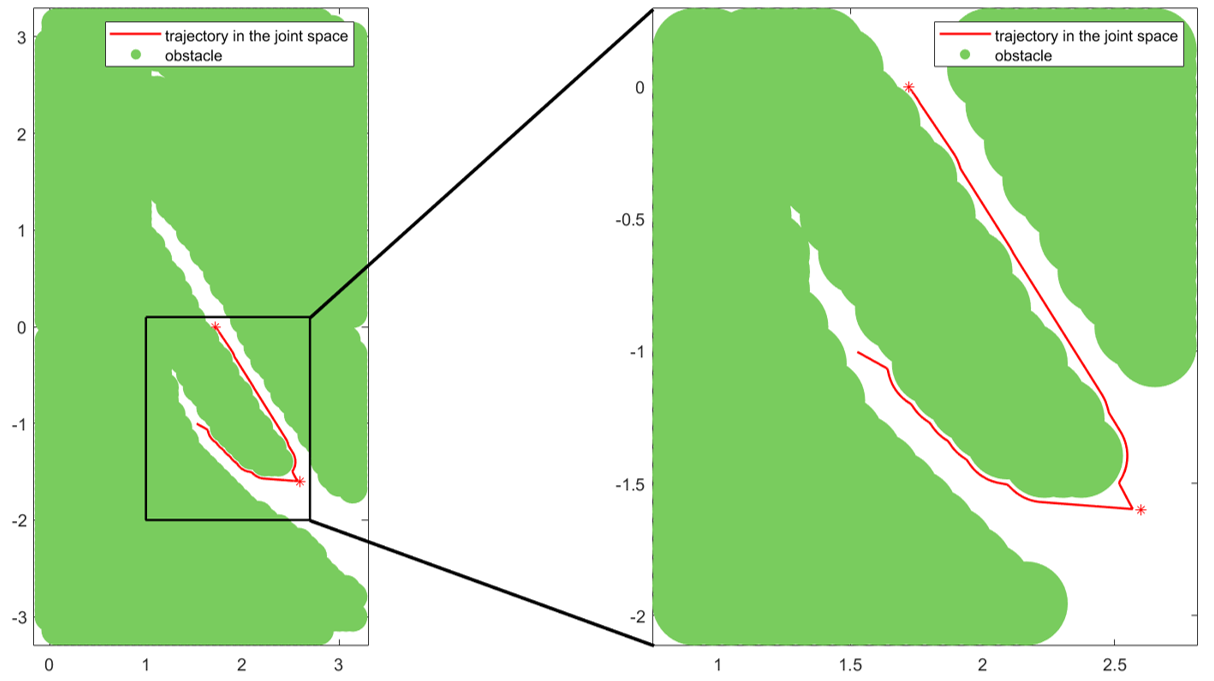}
	\caption{Unsafe regions and the motion of the $2$-link robotic manipulator in the joint space.}
	\label{figure.simulationjointspace}
\end{figure}

Figure \ref{figure.vstartsim} shows the trajectories of the velocity reference signal $v^*$ in the numerical simulation with the two different safety controllers. It can be observed that the velocity reference signals keep within the desired range, and coincide with each other.  One controller consists of \eqref{QPvirtualcontrolllaw1}--\eqref{QPvirtualcontrolllaw3} and \eqref{ic1}, and the other one consisits of \eqref{QPvirtualcontrolllaw1n}--\eqref{QPvirtualcontrolllaw2n} and \eqref{ic1}. Figures \ref{figure.simulationworkspace} and \ref{figure.simulationjointspace} show the motion of the manipulator with the first controller in the workspace and in the joint space, respectively. It can be observed that the motion of the manipulator keeps far away enough from the ball obstacles, as expected.

Since the velocity reference signals $v^*$ generated by different controllers are quite close to each other, the second controller can also guarantee the safety of the manipulator. The simulation result is not given here due to space limitation. The simulation results verify Theorems \ref{theorem.P1P2} and \ref{theorem.P2P3}.

Moreover, we test the computation times of the two controllers using different numbers of ball obstacles to represent the unsafe regions in the joint space. We use $N$ to represent the number of the ball obstacles. MATLAB-based numerical simulations are performed on a Windows laptop with an Intel Core i7-10510U CPU @ 1.80GHz. The computation time of the two controllers per iteration are given by Table \ref{comparesss}. It is shown that reducing the number of ball obstacles helps save computation time. 

\begin{table}[htbp] 
	\centering
	\caption{Computation Time of the Controllers (ms)}
	\tabcolsep=0.4cm 
	\label{comparesss} 
	\begin{tabular}{ccc}    
		\toprule  
		&\multicolumn{2}{c}{Type of controller}\\  
		N &\eqref{QPvirtualcontrolllaw1}--\eqref{QPvirtualcontrolllaw3}, \eqref{ic1}& \eqref{QPvirtualcontrolllaw1n}--\eqref{QPvirtualcontrolllaw2n}, \eqref{ic1} \\
		\midrule   
		3457& 15.5 & 6.7  \\
		8747& 38.9 & 6.9\\
		13617& 46.5 & 8.3  \\
		\bottomrule   
	\end{tabular}
\end{table}   

\subsection{Experimental Results}

\begin{figure}[h!]
	\centering
	\includegraphics[width=1\linewidth]{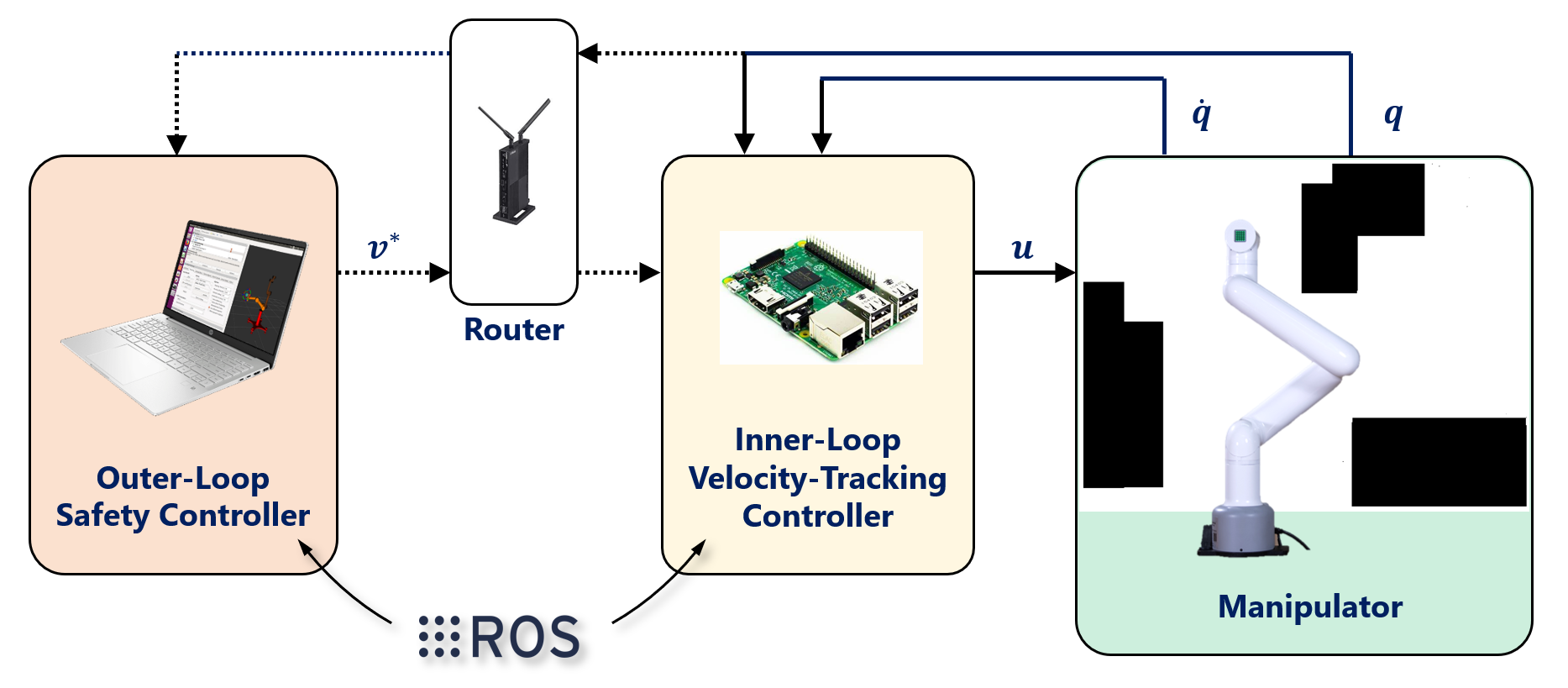}
	\caption{The experimental set-up.}
	\label{figure.experiment_connection}
\end{figure}

In this subsection, we verify our control methods based on the experimental system shown in Figure \ref{figure.experiment_connection}. The experimental system is composed of a $6$-DOF manipulator, a laptop and a wireless router. We use the second and third joints of the manipulator and lock the other four joints to verify the effectiveness of the proposed methods.

In our experiment setup, the QP-based safety-critical controllers in the form of \eqref{QPvirtualcontrolllaw1}--\eqref{QPvirtualcontrolllaw3} and \eqref{QPvirtualcontrolllaw1n}--\eqref{QPvirtualcontrolllaw2n} are implemented on the laptop, which communicates with the manipulator through a wireless local area network. Since the manipulator is already equipped with an inner-loop velocity-tracking controller, we only validate the outer-loop QP-based virtual control law. We perform system identification for the velocity tracking loop using Matlab System Identification Tool-box with $v^*$ as the input and $\dot{q}$ as the output. Then, we modify some controller parameters as follows:
\begin{align*}
	d_r&=0.0209,\hspace{20pt}\bar{v}\hspace{4pt}=0.5585,\hspace{20pt}\bar{v}_c=0.2094,\notag\\
	d_f&=0.0091,\hspace{20pt}d_h=0.000002,\notag\\
	\alpha_c(r)&=\begin{cases}
		1.8r,~&\text{for}~r\leq 0,\\
		61.5382r,~&\text{for}~0<r\leq 0.0091,\\
		1.8(r-0.0091)+0.5585,~&\text{for}~0.0091<r,
	\end{cases}\\
	\phi(r)&=\max\{-1.4881s+1.4828,0\},~r\in[-1,1].
\end{align*}

\begin{figure}[h!]
	\centering
	\includegraphics[width=1\linewidth]{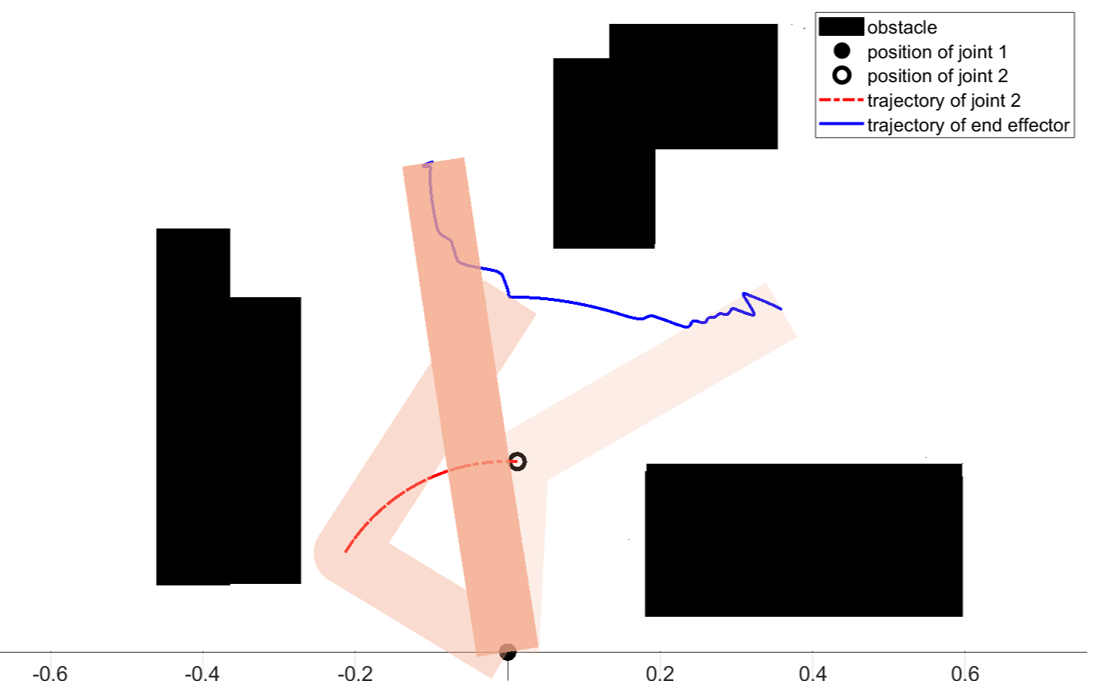}
	\caption{The motion of the $6$-link robotic manipulator in the workspace with the safety-critical controller \eqref{QPvirtualcontrolllaw1}--\eqref{QPvirtualcontrolllaw3}.}
	\label{figure.phyworkspace}
\end{figure}

\begin{figure}[h!]
	\centering
	\includegraphics[width=1\linewidth]{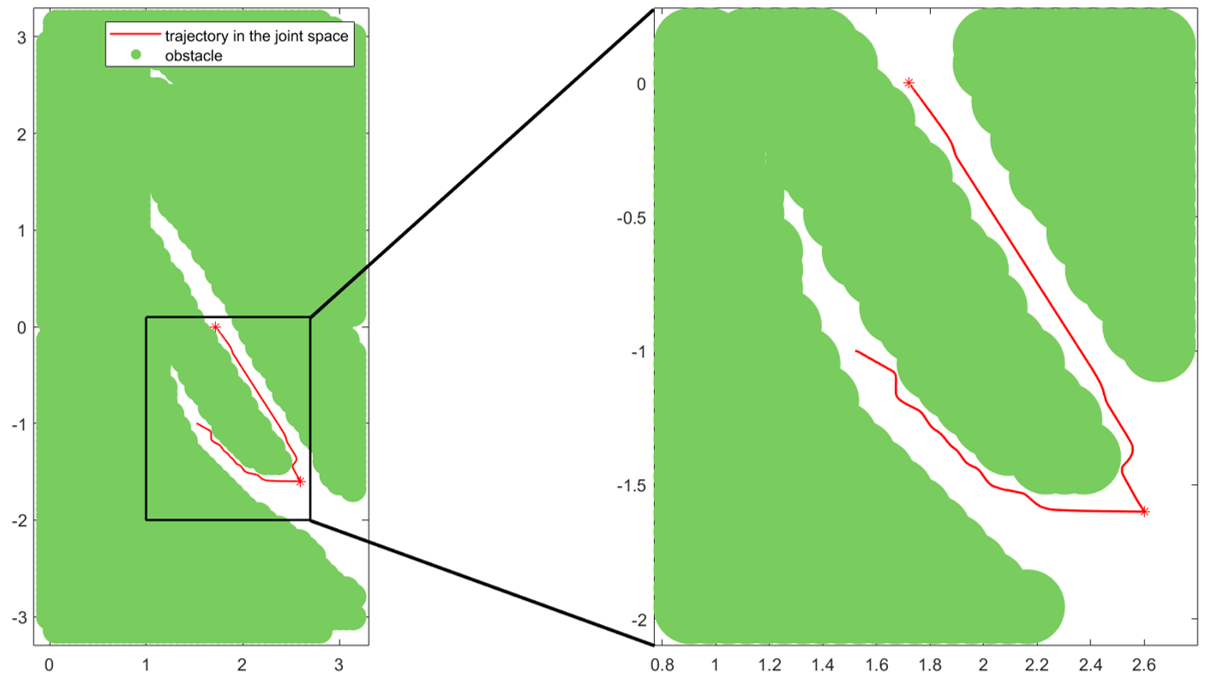}
	\caption{Unsafe regions and the motion of the $6$-link robotic manipulator in the joint space with the safety-critical controller \eqref{QPvirtualcontrolllaw1}--\eqref{QPvirtualcontrolllaw3}.}
	\label{figure.phyjointspace}
\end{figure}

\begin{figure}[h!]
	\centering
	\includegraphics[width=1\linewidth]{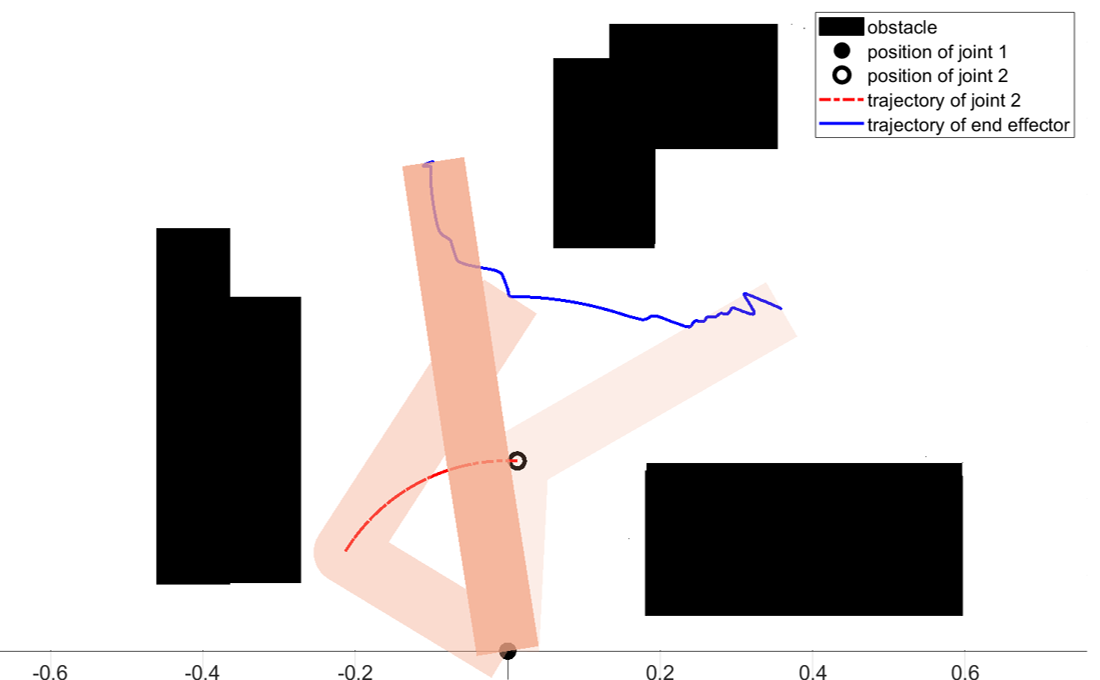}
	\caption{The motion of the $6$-link robotic manipulator in the workspace with the safety-critical controller \eqref{QPvirtualcontrolllaw1n}--\eqref{QPvirtualcontrolllaw2n}.}
	\label{figure.phyworkspace1}
\end{figure}

\begin{figure}[h!]
	\centering
	\includegraphics[width=1\linewidth]{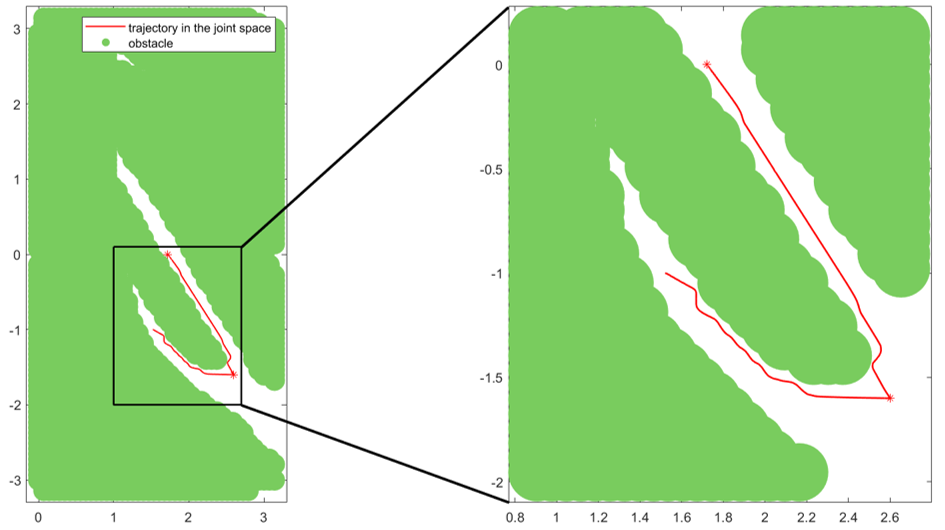}
	\caption{Unsafe regions and the motion of the $6$-link robotic manipulator in the joint space with the safety-critical controller \eqref{QPvirtualcontrolllaw1n}--\eqref{QPvirtualcontrolllaw2n}.}
	\label{figure.phyjointspace1}
\end{figure}

\begin{figure}[h!]
	\centering
	\includegraphics[width=1\linewidth]{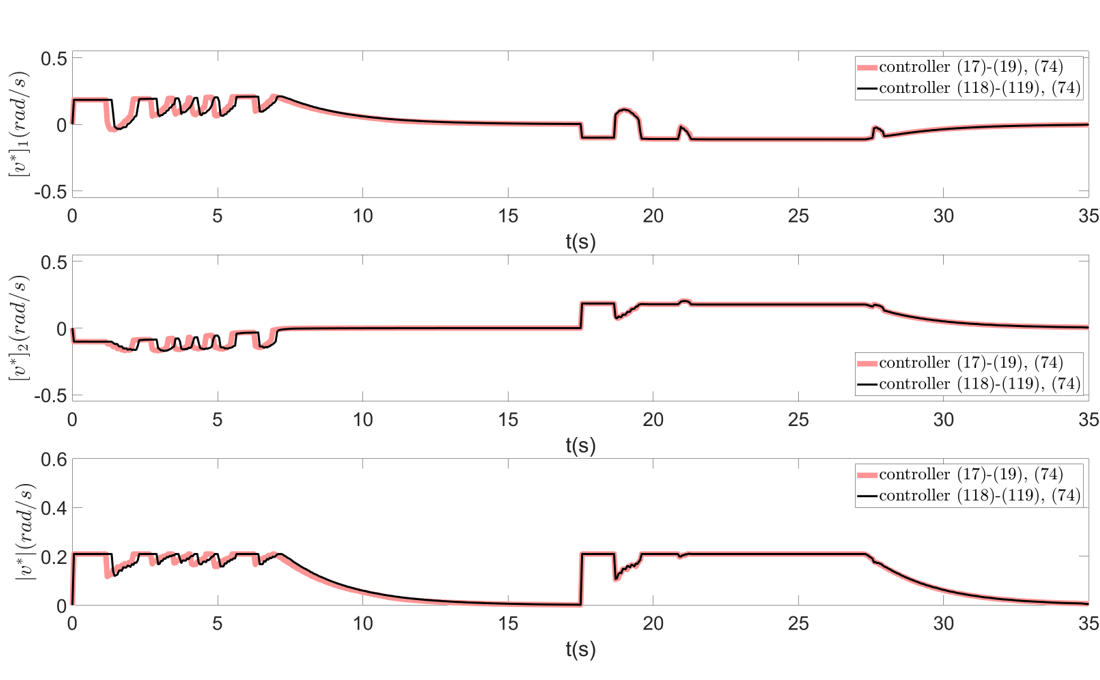}
	\caption{Trajectories of the velocity reference signals $v^*$ in the experiment.}
	\label{figure.vstarexperiment}
\end{figure}

Figures \ref{figure.phyworkspace} and \ref{figure.phyjointspace} show the motion of the manipulator in the workspace and the joint space with the proposed safety-critical controller \eqref{QPvirtualcontrolllaw1}--\eqref{QPvirtualcontrolllaw3}; Figures \ref{figure.phyworkspace1} and \ref{figure.phyjointspace1} show the motion of the manipulator in the workspace and the joint space with the proposed safety-critical controller \eqref{QPvirtualcontrolllaw1n}--\eqref{QPvirtualcontrolllaw2n}. Figure \ref{figure.vstarexperiment} shows the trajectories of the velocity reference signals $v^*$ generated by different controllers.

It can be observed that, with both controllers, the velocity reference signals keep within the desired range, and the motions of the manipulator keep far away enough from the ball obstacles, as expected.

\section{Conclusions}\label{conclusion}

This paper has proposed a systematic solution to the safety control of EL systems subject to multiple obstacles and velocity constraints. The proposed constraint-satisfaction controller comprises a refined QP-based outer-loop control law to handle the position constraints generated by multiple obstacles and a nonlinear inner-loop control law for velocity-tracking. The feasible set of the QP-based control law is reshaped based on an appropriately chosen positive basis to overcome the fundamental challenge caused by the possibly non-Lipschitz continuity of a standard QP-based algorithm when there are multiple constraints. The control objective of constraint-satisfaction is achievable if the ball obstacles describing the position constraints satisfy a mild distance condition. The main result is proved by constructing a max-type Lyapunov-like function for the closed-loop system.

The achievement in this paper would help solve safety-critical control problems for EL systems and other more general robotic systems working in time-varying, dynamic and interactive environments.

\bibliographystyle{IEEEtran}        

\bibliography{Autoreference}           

\begin{thebibliography}{}
\providecommand{\url}[1]{#1}
\csname url@samestyle\endcsname
\providecommand{\newblock}{\relax}
\providecommand{\bibinfo}[2]{#2}
\providecommand{\BIBentrySTDinterwordspacing}{\spaceskip=0pt\relax}
\providecommand{\BIBentryALTinterwordstretchfactor}{4}
\providecommand{\BIBentryALTinterwordspacing}{\spaceskip=\fontdimen2\font plus
\BIBentryALTinterwordstretchfactor\fontdimen3\font minus
  \fontdimen4\font\relax}
\providecommand{\BIBforeignlanguage}[2]{{%
\expandafter\ifx\csname l@#1\endcsname\relax
\typeout{** WARNING: IEEEtran.bst: No hyphenation pattern has been}%
\typeout{** loaded for the language `#1'. Using the pattern for}%
\typeout{** the default language instead.}%
\else
\language=\csname l@#1\endcsname
\fi
#2}}
\providecommand{\BIBdecl}{\relax}
\BIBdecl

\end{thebibliography}


\begin{thebibliography}{10}
\providecommand{\url}[1]{#1}
\csname url@samestyle\endcsname
\providecommand{\newblock}{\relax}
\providecommand{\bibinfo}[2]{#2}
\providecommand{\BIBentrySTDinterwordspacing}{\spaceskip=0pt\relax}
\providecommand{\BIBentryALTinterwordstretchfactor}{4}
\providecommand{\BIBentryALTinterwordspacing}{\spaceskip=\fontdimen2\font plus
\BIBentryALTinterwordstretchfactor\fontdimen3\font minus
  \fontdimen4\font\relax}
\providecommand{\BIBforeignlanguage}[2]{{%
\expandafter\ifx\csname l@#1\endcsname\relax
\typeout{** WARNING: IEEEtran.bst: No hyphenation pattern has been}%
\typeout{** loaded for the language `#1'. Using the pattern for}%
\typeout{** the default language instead.}%
\else
\language=\csname l@#1\endcsname
\fi
#2}}
\providecommand{\BIBdecl}{\relax}
\BIBdecl

\bibitem{Latombe-book-1991}
J.-C. Latombe, \emph{Robot Motion Planning}.\hskip 1em plus 0.5em minus
  0.4em\relax Springer, 1991.

\bibitem{Arkin-book-1998}
R.~C. Arkin, \emph{Behavior-Based Robotics}.\hskip 1em plus 0.5em minus
  0.4em\relax MIT Press, 1998.

\bibitem{Choset-book-2005}
H.~Choset, K.~M. Lynch, S.~Hutchinson, G.~A. Kantor, W.~Burgard, L.~E. Kavraki,
  and S.~Thrun, \emph{Principles of Robot Motion: Theory, Algorithms, and
  Implementations}.\hskip 1em plus 0.5em minus 0.4em\relax MIT Press, 2005.

\bibitem{Ren-Beard-book-2008}
W.~Ren and R.~W. Beard, \emph{Distributed Consensus in Multi-vehicle
  Cooperative Control: Theory and Applications}.\hskip 1em plus 0.5em minus
  0.4em\relax Springer, 2008.

\bibitem{Bullo-Cortes-Martinez-book-2009}
F.~Bullo, J.~Cort{\'{e}}s, and S.~Martinez, \emph{Distributed Control of
  Robotic Networks: A Mathematical Approach to Motion Coordination
  Algorithms}.\hskip 1em plus 0.5em minus 0.4em\relax Princeton University
  Press, 2009.

\bibitem{Mesbahi-Egerstedt-book-2010}
M.~Mesbahi and M.~Egerstedt, \emph{Graph Theoretic Methods in Multiagent
  Networks}.\hskip 1em plus 0.5em minus 0.4em\relax Princeton University Press,
  2010.

\bibitem{Lynch-Park-2017-Book}
K.~M. Lynch and F.~C. Park, \emph{Modern Robotics: Mechanics, Planning, and
  Control}.\hskip 1em plus 0.5em minus 0.4em\relax Cambridge University Press,
  2017.

\bibitem{Slotine-Siciliano-1991-ICAR}
S.~B. Slotine and B.~Siciliano, ``A general framework for managing multiple
  tasks in highly redundant robotic systems,'' in \emph{Proceedings of the 5th
  International Conference on Advanced Robotics}, vol.~2, 1991, pp. 1211--1216.

\bibitem{Polak-Yang-Mayne-SIAMControl-1993}
E.~Polak, T.~H. Yang, and D.~Q. Mayne, ``A method of centers based on barrier
  functions for solving optimal control problems with continuum state and
  control constraints,'' \emph{SIAM Journal on Control and Optimization},
  vol.~31, pp. 159--179, 1993.

\bibitem{Wills-Heath-Auto-2004}
A.~G. Wills and W.~P. Heath, ``Barrier function based model predictive
  control,'' \emph{Automatica}, vol.~40, pp. 1415--1422, 2004.

\bibitem{Ngo-Mahony-Jiang-CDC-2005}
K.~B. Ngo, R.~Mahony, and Z.~P. Jiang, ``Integrator backstepping using barrier
  functions for systems with multiple state constraints,'' in \emph{Proceedings
  of the 44th IEEE Conference on Decision and Control}, 2005, pp. 8306--8312.

\bibitem{Tee-Ge-Tay-Auto-2009}
K.~P. Tee, S.~S. Ge, and E.~H. Tay, ``Barrier {L}yapunov functions for the
  control of output-constrained nonlinear systems,'' \emph{Automatica},
  vol.~45, pp. 918--927, 2009.

\bibitem{Wieland-Allgower-NOLCOS-2007}
P.~Wieland and F.~Allgower, ``Constructive safety using control barrier
  functions,'' in \emph{Proceedings of the 7th IFAC Symposium on Nonlinear
  Control System}, 2007, pp. 462--467.

\bibitem{Prajna-Jadbabaie-Pappas-TAC-2007}
S.~Prajna, A.~Jadbabaie, and G.~J. Pappas, ``A framework for worst-case and
  stochastic safety verification using barrier certificates,'' \emph{IEEE
  Transactions on Automatic Control}, vol.~52, pp. 1415--1428, 2007.

\bibitem{Ames-Grizzle-Tabuada-CDC-2014}
A.~D. Ames, J.~W. Grizzle, and P.~Tabuada, ``Control barrier function based
  quadratic programs with application to adaptive cruise control,'' in
  \emph{Proceedings of the 53rd IEEE Conference on Decision and Control}, 2014,
  pp. 6271--6278.

\bibitem{Wisniewski-Sloth-TAC-2016}
R.~Wisniewski and C.~Sloth, ``Converse barrier certificate theorems,''
  \emph{IEEE Transactions on Automatic Control}, vol.~61, pp. 1356--1361, 2016.

\bibitem{Romdlony-Jayawardhana-Auto-2016}
M.~Z. Romdlony and B.~Jayawardhana, ``Stabilization with guaranteed safety
  using control {L}yapunov–barrier function,'' \emph{Automatica}, vol.~66,
  pp. 39--47, 2016.

\bibitem{Artstein-NA-1983}
Z.~Artstein, ``Stabilization with relaxed controls,'' \emph{Nonlinear Analysis:
  Theory, Methods \& Applications}, vol.~7, pp. 1163--1173, 1983.

\bibitem{Ames-Xu-Grizzle-Tabuada-TAC-2017}
A.~D. Ames, X.~Xu, J.~W. Grizzle, and P.~Tabuada, ``Control barrier function
  based quadratic programs for safety critical systems,'' \emph{IEEE
  Transactions on Automatic Control}, vol.~62, pp. 3861--3876, 2017.

\bibitem{Xu-Tabuada-Grizzle-Ames-IFAC-2015}
X.~Xu, P.~Tabuada, J.~W. Grizzle, and A.~D. Ames, ``Robustness of control
  barrier functions for safety critical control,'' in \emph{Proceedings of the
  19th IFAC World Congress}, 2014, pp. 054--061.

\bibitem{Jankovic-Auto-2018}
M.~Jankovic, ``Robust control barrier functions for constrained stabilization
  of nonlinear systems,'' \emph{Automatica}, vol.~96, pp. 359--367, 2018.

\bibitem{Xiao-TAC-2021}
W.~Xiao and C.~Belta, ``High order control barrier functions,'' \emph{IEEE
  Transactions on Automatic Control}, vol.~67, no.~7, pp. 3655--3662, 2021.

\bibitem{Nakamura-1990-Book}
Y.~Nakamura, \emph{Advanced Robotics: Redundancy and Optimization}.\hskip 1em
  plus 0.5em minus 0.4em\relax Addison-Wesley, 1990.

\bibitem{Escande-Mansard-Wieber-IJRR-2014}
A.~Escande, N.~Mansard, and P.-B. Wieber, ``Hierarchical quadratic programming:
  Fast online humanoid-robot motion generation,'' \emph{International Journal
  of Robotics Research}, vol.~33, pp. 1006--1028, 2014.

\bibitem{Mellinger-Kumar-ICRA-2011}
D.~Mellinger and V.~Kumar, ``Minimum snap trajectory generation and control for
  quadrotors,'' in \emph{Proceedings of the 2011 IEEE International Conference
  on Robotics and Automation}, 2011, pp. 2520--2525.

\bibitem{Ames-Powell-CPS-2013}
A.~D. Ames and M.~Powell, ``Towards the unification of locomotion and
  manipulation through control {L}yapunov functions and quadratic programs,''
  in \emph{Control of Cyber-Physical Systems}, D.~Tarraf, Ed.\hskip 1em plus
  0.5em minus 0.4em\relax Springer, 2013, pp. 219--240.

\bibitem{Glotfelter-Paul-TAC-2021}
P.~Glotfelter, J.~Cortés, and M.~Egerstedt, ``A nonsmooth approach to
  controller synthesis for boolean specifications,'' \emph{IEEE Transactions on
  Automatic Control}, vol.~66, pp. 5160--5174, 2021.

\bibitem{Wang-Ames-Egerstedt-TRO-2017}
L.~Wang, A.~D. Ames, and M.~Egerstedt, ``Safety barrier certificates for
  collisions-free multirobot systems,'' \emph{IEEE Transactions on Robotics},
  vol.~33, pp. 661--674, 2017.

\bibitem{Wilson-Egerstedt-CSM-2020}
S.~Wilson, P.~Glotfelter, L.~Wang, S.~Mayya, G.~Notomista, M.~Mote, and
  M.~Egerstedt, ``The {Robotarium}: Globally impactful opportunities,
  challenges, and lessons learned in remote-access, distributed control of
  multirobot systems,'' \emph{IEEE Control Systems Magazine}, vol.~40, pp.
  26--44, 2020.

\bibitem{Cortez-Oetomo-Manzie-Choong-TCST-2019}
W.~Shaw~Cortez, D.~Oetomo, C.~Manzie, and P.~Choong, ``Control barrier
  functions for mechanical systems: Theory and application to robotic
  grasping,'' \emph{IEEE Transactions on Control Systems Technology}, vol.~29,
  pp. 530--545, 2021.

\bibitem{Singletary-Guffey-RAL-2022}
A.~Singletary, W.~Guffey, T.~G. Molnar, R.~Sinnet, and A.~D. Ames,
  ``Safety-critical manipulation for collision-free food preparation,''
  \emph{IEEE Robotics and Automation Letters}, vol.~7, pp. 10\,954--10\,961,
  2022.

\bibitem{Xiong-Ames-RAL-2021}
X.~Xiong and A.~Ames, ``Slip walking over rough terrain via h-lip stepping and
  backstepping-barrier function inspired quadratic program,'' \emph{IEEE
  Robotics and Automation Letters}, vol.~6, pp. 2122--2129, 2021.

\bibitem{Wu-Li-Kan-Gao-TC-2019}
X.~Wu, Z.~Li, Z.~Kan, and H.~Gao, ``Reference trajectory reshaping optimization
  and control of robotic exoskeletons for human-robot co-manipulation,''
  \emph{IEEE Transactions on Cybernetics}, vol.~50, no.~8, pp. 3740--3751,
  2019.

\bibitem{Singletary-Nilsson-Gurriet-Ames-IROS-2019}
A.~Singletary, P.~Nilsson, T.~Gurriet, and A.~D. Ames, ``Online active safety
  for robotic manipulators,'' in \emph{Proceedings of 2019 IEEE/RSJ
  International Conference on Intelligent Robots and Systems}, 2019, pp.
  173--178.

\bibitem{Lozamo-TSMC-1981}
T.~Lozano-Perez, ``Automatic planning of manipulator transfer movements,''
  \emph{IEEE Transactions on Systems, Man, and Cybernetics}, vol.~11, no.~10,
  pp. 681--698, 1981.

\bibitem{Dietrich-RAM-2012}
A.~Dietrich, T.~Wimbock, A.~Albu-Sch{\"a}ffer, and G.~Hirzinger, ``Reactive
  whole-body control: Dynamic mobile manipulation using a large number of
  actuated degrees of freedom,'' \emph{IEEE Robotics \& Automation Magazine},
  vol.~19, no.~2, pp. 20--33, 2012.

\bibitem{Youakim-RAM-2017}
D.~Youakim, P.~Ridao, N.~Palomeras, F.~Spadafora, D.~Ribas, and M.~Muzzupappa,
  ``Moveit!: Autonomous underwater free-floating manipulation,'' \emph{IEEE
  Robotics \& Automation Magazine}, vol.~24, no.~3, pp. 41--51, 2017.

\bibitem{Khatib-JRR-1986}
O.~Khatib, ``Real-time obstacle avoidance for manipulators and mobile robots,''
  \emph{International Journal of Robotics Research}, vol.~5, pp. 90--98, 1986.

\bibitem{Singletary-Kolathaya-CSL-2022}
A.~Singletary, S.~Kolathaya, and A.~D. Ames, ``Safety-critical kinematic
  control of robotic systems,'' \emph{IEEE Control Systems Letters}, vol.~6,
  pp. 139--144, 2022.

\bibitem{Cortez-Dimarogonas-Auto-2022}
W.~Shaw~Cortez and D.~V. Dimarogonas, ``Safe-by-design control for
  {Euler}–{Lagrange} systems,'' \emph{Automatica}, vol. 146, no. 110620,
  2022.

\bibitem{Krstic-Kanellakopoulos-Kokotovic-1995-Book}
M.~Krstic, I.~Kanellakopoulos, and P.~V. Kokotovic, \emph{Nonlinear and
  Adaptive Control Design}.\hskip 1em plus 0.5em minus 0.4em\relax Wiley, 1995.

\bibitem{Hager-SIAMControl-1979}
W.~W. Hager, ``Lipschitz continuity for constrained processes,'' \emph{SIAM
  Journal on Control and Optimization}, vol.~17, pp. 321--338, 1979.

\bibitem{Wu-Liu-Niu-Jiang-RAL-2022}
S.~Wu, T.~Liu, Q.~Niu, and Z.~P. Jiang, ``Continuous safety control of mobile
  robots in cluttered environments,'' \emph{IEEE Robotics and Automation
  Letter}, vol.~7, no.~3, pp. 8012--8019, 2022.

\bibitem{Wu-Liu-Egerstedt-Jiang-TAC-2023}
S.~Wu, T.~Liu, M.~Egerstedt, and Z.~P. Jiang, ``Quadratic programming for
  continuous control of safety-critical multi-agent systems under
  uncertainty,'' \emph{IEEE Transactions on Automatic Control}, vol.~68,
  no.~11, pp. 6664--6679, 2023.

\bibitem{Cortez-CSL-2022}
W.~Shaw~Cortez, X.~Tan, and D.~V. Dimarogonas, ``A robust, multiple control
  barrier function framework for input constrained systems,'' \emph{IEEE
  Control Systems Letters}, vol.~6, pp. 1742--1747, 2022.

\bibitem{Khalil-book-2002}
H.~K. Khalil, \emph{Nonlinear Systems}, 3rd~ed.\hskip 1em plus 0.5em minus
  0.4em\relax NJ: Prentice-Hall, 2002.

\bibitem{Murray-Li-Sastry-1994-book}
R.~M. Murray, Z.~Li, and S.~S. Sastry, \emph{A Mathematical Introduction to
  Robotic Manipulation}.\hskip 1em plus 0.5em minus 0.4em\relax CRC Press,
  1994.

\bibitem{Craig-book-2005}
J.~J. Craig, \emph{Introduction to Robotics: Mechanics and Control}, 3rd~ed.,
  2005.

\bibitem{Bertsekas-book-1997}
D.~P. Bertsekas, \emph{Nonlinear Programming}, 2nd~ed.\hskip 1em plus 0.5em
  minus 0.4em\relax Athena Scientific, 1999.

\bibitem{Cobzas-Miculescu-Nicolae-book-2019}
{\c{S}}.~Cobza{\c{s}}, R.~Miculescu, and A.~Nicolae, \emph{Lipschitz
  Functions}.\hskip 1em plus 0.5em minus 0.4em\relax Springer, 2019.

\bibitem{Clarke-book-1990}
F.~H. Clarke, \emph{Optimization and Nonsmooth Analysis}.\hskip 1em plus 0.5em
  minus 0.4em\relax SIAM, 1990.

\bibitem{Grunbaum-book-1967}
B.~Gr{\"u}nbaum, V.~Klee, M.~A. Perles, and G.~C. Shephard, \emph{Convex
  Polytopes}.\hskip 1em plus 0.5em minus 0.4em\relax Springer, 1967, vol.~16.

\bibitem{Della-2020-nonsmooth}
D.~R. Matteo, ``Non-smooth lyapunov functions for stability analysis of hybrid
  systems,'' Ph.D. dissertation, Institut National des Sciences Appliqu{\'e}es
  de Toulouse, 2020.

\bibitem{Lakshmikantham-book-1969}
V.~Lakshmikantham and S.~Leela, \emph{Differential and integral inequalities:
  theory and applications: volume I: ordinary differential equations}.\hskip
  1em plus 0.5em minus 0.4em\relax Academic press, 1969.

\bibitem{Jiang-Mareels-Wang-Auto-1996}
Z.~P. Jiang, I.~M. Mareels, and Y.~Wang, ``A {Lyapunov} formulation of the
  nonlinear small-gain theorem for interconnected iss systems,''
  \emph{Automatica}, vol.~32, no.~8, pp. 1211--1215, 1996.

\bibitem{Mahil-Ahmed-MWSCAS-2016}
S.~M. Mahil and A.~A. Durra, ``Modeling analysis and simulation of 2-{DOF}
  robotic manipulator,'' in \emph{2016 IEEE 59th International Midwest
  Symposium on Circuits and Systems}, 2016, pp. 1--4.

\end{thebibliography}


\begin{thebibliography}{10}

\bibitem{Ames-Grizzle-Tabuada-CDC-2014}
A.~D. Ames, J.~W. Grizzle, and P.~Tabuada.
\newblock Control barrier function based quadratic programs with application to
  adaptive cruise control.
\newblock In {\em Proceedings of the 53rd IEEE Conference on Decision and
  Control}, pages 6271--6278, 2014.

\bibitem{Ames-Xu-Grizzle-Tabuada-TAC-2017}
A.~D. Ames, X.~Xu, J.~W. Grizzle, and P.~Tabuada.
\newblock Control barrier function based quadratic programs for safety critical
  systems.
\newblock {\em IEEE Transactions on Automatic Control}, 62:3861--3876, 2017.

\bibitem{Ames-Powell-CPS-2013}
Aaron~D. Ames and Matthew Powell.
\newblock Towards the unification of locomotion and manipulation through
  control {L}yapunov functions and quadratic programs.
\newblock In D.~Tarraf, editor, {\em Control of Cyber-Physical Systems}, pages
  219--240. Springer, 2013.

\bibitem{Arkin-book-1998}
R.~C. Arkin.
\newblock {\em Behavior-Based Robotics}.
\newblock MIT Press, 1998.

\bibitem{Artstein-NA-1983}
Zvi Artstein.
\newblock Stabilization with relaxed controls.
\newblock {\em Nonlinear Analysis: Theory, Methods \& Applications},
  7:1163--1173, 1983.

\bibitem{Bertsekas-book-1997}
D.~P. Bertsekas.
\newblock {\em Nonlinear Programming}.
\newblock Athena Scientific, second edition, 1999.

\bibitem{Bullo-Cortes-Martinez-book-2009}
F.~Bullo, J.~Cort{\'{e}}s, and S.~Martinez.
\newblock {\em Distributed Control of Robotic Networks: A Mathematical Approach
  to Motion Coordination Algorithms}.
\newblock Princeton University Press, 2009.

\bibitem{Choset-book-2005}
Howie Choset, Kevin~M. Lynch, Seth Hutchinson, George~A. Kantor, Wolfram
  Burgard, Lydia~E. Kavraki, and Sebastian Thrun.
\newblock {\em Principles of Robot Motion: Theory, Algorithms, and
  Implementations}.
\newblock MIT Press, 2005.

\bibitem{Cortez-Oetomo-Manzie-Choong-TCST-2019}
W.~S. Cortez, D.~Oetomo, C.~Manzie, and P.~Choong.
\newblock Control barrier functions for mechanical systems: Theory and
  application to robotic grasping.
\newblock {\em IEEE Transactions on Control Systems Technology}, 29:530--545,
  2021.

\bibitem{Cortez-Dimarogonas-ACC-2020}
Wenceslao~Shaw Cortez and Dimos~V. Dimarogonas.
\newblock Correct-by-design control barrier functions for euler-lagrange
  systems with input constraints.
\newblock In {\em 2020 American Control Conference (ACC)}, pages 950--955,
  2020.

\bibitem{Craig-book-2005}
John~J. Craig.
\newblock {\em Introduction to Robotics: Mechanics and Control}.
\newblock 3rd edition, 2005.

\bibitem{Escande-Mansard-Wieber-IJRR-2014}
Adrien Escande, Nicolas Mansard, and Pierre-Brice Wieber.
\newblock Hierarchical quadratic programming: Fast online humanoid-robot motion
  generation.
\newblock {\em International Journal of Robotics Research}, 33:1006--1028,
  2014.

\bibitem{Giorgi-1992-Book}
Giorgio Giorgi and S{\'a}ndor Koml{\'o}si.
\newblock {\em Dini derivatives in optimization—Part I}.
\newblock Springer, 1992.

\bibitem{Glotfelter-Paul-TAC-2021}
Paul Glotfelter, Jorge Cortés, and Magnus Egerstedt.
\newblock A nonsmooth approach to controller synthesis for boolean
  specifications.
\newblock {\em IEEE Transactions on Automatic Control}, 66:5160--5174, 2021.

\bibitem{Hager-SIAMControl-1979}
William~W. Hager.
\newblock Lipschitz continuity for constrained processes.
\newblock {\em SIAM Journal on Control and Optimization}, 17:321--338, 1979.

\bibitem{Jankovic-Auto-2018}
Mrdjan Jankovic.
\newblock Robust control barrier functions for constrained stabilization of
  nonlinear systems.
\newblock {\em Automatica}, 96:359--367, 2018.

\bibitem{Jiang-Mareels-Wang-Auto-1996}
Zhong-Ping Jiang, Iven~MY Mareels, and Yuan Wang.
\newblock A {Lyapunov} formulation of the nonlinear small-gain theorem for
  interconnected iss systems.
\newblock {\em Automatica}, 32(8):1211--1215, 1996.

\bibitem{Khalil-book-2002}
H.~K. Khalil.
\newblock {\em Nonlinear Systems}.
\newblock NJ: Prentice-Hall, third edition, 2002.

\bibitem{Kim-JRNAL-2017}
Jinho Kim, Kevin Chang, Brian Schwarz, Andrew~S. Lee, S.~Andrew Gadsden, and
  Mohammad Al-Shabi.
\newblock Dynamic model and motion control of a robotic manipulator.
\newblock {\em Journal of Robotics, Networking and Artificial Life},
  4(2):138--141, 2017.

\bibitem{Krstic-Kanellakopoulos-Kokotovic-1995-Book}
Miroslav Krstic, Ioannis Kanellakopoulos, and Petar~V. Kokotovic.
\newblock {\em Nonlinear and Adaptive Control Design}.
\newblock Wiley, 1995.

\bibitem{Latombe-book-1991}
J.-C. Latombe.
\newblock {\em Robot Motion Planning}.
\newblock Springer, 1991.

\bibitem{Lynch-Park-2017-Book}
Kevin~M Lynch and Frank~C Park.
\newblock {\em Modern Robotics: Mechanics, Planning, and Control}.
\newblock Cambridge University Press, 2017.

\bibitem{Mellinger-Kumar-ICRA-2011}
Daniel Mellinger and Vijay Kumar.
\newblock Minimum snap trajectory generation and control for quadrotors.
\newblock In {\em Proceedings of the 2011 IEEE International Conference on
  Robotics and Automation}, pages 2520--2525, 2011.

\bibitem{Mesbahi-Egerstedt-book-2010}
M.~Mesbahi and M.~Egerstedt.
\newblock {\em Graph Theoretic Methods in Multiagent Networks}.
\newblock Princeton University Press, 2010.

\bibitem{Murray-Li-Sastry-1994-book}
Richard~M. Murray, Zexiang Li, and S.~Shankar Sastry.
\newblock {\em A Mathematical Introduction to Robotic Manipulation}.
\newblock CRC Press, 1994.

\bibitem{Nakamura-1990-Book}
Yoshihiko Nakamura.
\newblock {\em Advanced Robotics: Redundancy and Optimization}.
\newblock Addison-Wesley Longman Publishing Co., Inc., 1990.

\bibitem{Ngo-Mahony-Jiang-CDC-2005}
Khoi~B. Ngo, Robert Mahony, and Z.~P. Jiang.
\newblock Integrator backstepping using barrier functions for systems with
  multiple state constraints.
\newblock In {\em Proceedings of the 44th IEEE Conference on Decision and
  Control}, pages 8306--8312, 2005.

\bibitem{Polak-Yang-Mayne-SIAMControl-1993}
E.~Polak, T.~H. Yang, and D.~Q. Mayne.
\newblock A method of centers based on barrier functions for solving optimal
  control problems with continuum state and control constraints.
\newblock {\em SIAM Journal on Control and Optimization}, 31:159--179, 1993.

\bibitem{Prajna-Jadbabaie-Pappas-TAC-2007}
S.~Prajna, A.~Jadbabaie, and G.~J. Pappas.
\newblock A framework for worst-case and stochastic safety verification using
  barrier certificates.
\newblock {\em IEEE Transactions on Automatic Control}, 52:1415--1428, 2007.

\bibitem{Ren-Beard-book-2008}
W.~Ren and R.~W. Beard.
\newblock {\em Distributed Consensus in Multi-vehicle Cooperative Control:
  Theory and Applications}.
\newblock Springer, 2008.

\bibitem{Romdlony-Jayawardhana-Auto-2016}
M.~Z. Romdlony and B.~Jayawardhana.
\newblock Stabilization with guaranteed safety using control
  {L}yapunov–barrier function.
\newblock {\em Automatica}, 66:39--47, 2016.

\bibitem{Singletary-Guffey-RAL-2022}
Andrew Singletary, William Guffey, Tamas~G. Molnar, Ryan Sinnet, and Aaron~D.
  Ames.
\newblock Safety-critical manipulation for collision-free food preparation.
\newblock {\em IEEE Robotics and Automation Letters}, 7:10954--10961, 2022.

\bibitem{Singletary-Kolathaya-CSL-2022}
Andrew Singletary, Shishir Kolathaya, and Aaron~D. Ames.
\newblock Safety-critical kinematic control of robotic systems.
\newblock {\em IEEE Control Systems Letters}, 6:139--144, 2022.

\bibitem{Singletary-Nilsson-Gurriet-Ames-IROS-2019}
Andrew Singletary, Petter Nilsson, Thomas Gurriet, and Aaron~D Ames.
\newblock Online active safety for robotic manipulators.
\newblock In {\em 2019 IEEE/RSJ International Conference on Intelligent Robots
  and Systems (IROS)}, pages 173--178. IEEE, 2019.

\bibitem{Slotine-Siciliano-1991-ICAR}
Siciliano~B Slotine and B~Siciliano.
\newblock A general framework for managing multiple tasks in highly redundant
  robotic systems.
\newblock In {\em Proceeding of the 5th International Conference on Advanced
  Robotics}, volume~2, pages 1211--1216, 1991.

\bibitem{Tee-Ge-Tay-Auto-2009}
K.~P. Tee, S.~S. Ge, and E.~H. Tay.
\newblock Barrier {L}yapunov functions for the control of output-constrained
  nonlinear systems.
\newblock {\em Automatica}, 45:918--927, 2009.

\bibitem{Wang-Ames-Egerstedt-TRO-2017}
L.~Wang, A.~D. Ames, and M.~Egerstedt.
\newblock Safety barrier certificates for collisions-free multirobot systems.
\newblock {\em IEEE Transactions on Robotics}, 33:661--674, 2017.

\bibitem{Wieland-Allgower-NOLCOS-2007}
P.~Wieland and F.~Allgower.
\newblock Constructive safety using control barrier functions.
\newblock In {\em Proceedings of the 7th IFAC Symposium on Nonlinear Control
  System}, pages 462--467, 2007.

\bibitem{Wills-Heath-Auto-2004}
Adrian~G. Wills and William~P. Heath.
\newblock Barrier function based model predictive control.
\newblock {\em Automatica}, 40:1415--1422, 2004.

\bibitem{Wilson-Egerstedt-CSM-2020}
Sean Wilson, Paul Glotfelter, Li~Wang, Siddharth Mayya, Gennaro Notomista, Mark
  Mote, and Magnus Egerstedt.
\newblock The robotarium: Globally impactful opportunities, challenges, and
  lessons learned in remote-access, distributed control of multirobot systems.
\newblock {\em IEEE Control Systems Magazine}, 40:26--44, 2020.

\bibitem{Wisniewski-Sloth-TAC-2016}
R.~Wisniewski and C.~Sloth.
\newblock Converse barrier certificate theorems.
\newblock {\em IEEE Transactions on Automatic Control}, 61:1356--1361, 2016.

\bibitem{Wu-Liu-Niu-Jiang-RAL-2022}
S.~Wu, T.~Liu, Q.~Niu, and Z.~P. Jiang.
\newblock Continuous safety control of mobile robots in cluttered environments.
\newblock {\em IEEE Robotics and Automation Letters}, 2022.

\bibitem{Wu-Liu-JSSC-2022}
Si~Wu and Tengfei Liu.
\newblock Safety control of a class of fully actuated systems subject to
  uncertain actuation dynamics.
\newblock {\em Journal of Systems Science and Complexity}, 35(2):543--558,
  2022.

\bibitem{Wu-Liu-Egerstedt-Jiang-TAC-2023}
Si~Wu, Tengfei Liu, Magnus Egerstedt, and Zhong-Ping Jiang.
\newblock Quadratic programming for continuous control of safety-critical
  multi-agent systems under uncertainty.
\newblock {\em Conditionally Accepted as Paper by IEEE Transactions on
  Automatic Control}, 2023.

\bibitem{Wu-Li-Kan-Gao-TC-2019}
Xiaoyu Wu, Zhijun Li, Zhen Kan, and Hongbo Gao.
\newblock Reference trajectory reshaping optimization and control of robotic
  exoskeletons for human-robot co-manipulation.
\newblock {\em IEEE Transactions on Cybernetics}, 50(8):3740--3751, 2019.

\bibitem{Xiao-TAC-2021}
Wei Xiao and Calin Belta.
\newblock High order control barrier functions.
\newblock {\em IEEE Transactions on Automatic Control}, 67(7):3655--3662, 2021.

\bibitem{Xiong-Ames-RAL-2021}
Xiaobin Xiong and Aaron Ames.
\newblock Slip walking over rough terrain via h-lip stepping and
  backstepping-barrier function inspired quadratic program.
\newblock {\em IEEE Robotics and Automation Letters}, 6:2122--2129, 2021.

\bibitem{Xu-Tabuada-Grizzle-Ames-IFAC-2015}
X.~Xu, P.~Tabuada, J.~W. Grizzle, and A.~D. Ames.
\newblock Robustness of control barrier functions for safety critical control.
\newblock In {\em Proceedings of the 19th IFAC World Congress}, pages 054--061,
  2014.

\end{thebibliography}

\end{document}